\pdfoutput=1

\documentclass[12pt]{article}
\usepackage{amsmath}
\usepackage{amsthm}
\usepackage{amssymb}
\usepackage{graphicx,psfrag}

\usepackage{enumerate}
\usepackage{natbib}
\usepackage{multirow}
\usepackage{float}
\usepackage{bm}
\usepackage{caption}
\usepackage{subcaption}
\usepackage{color}
\usepackage[margin=1in]{geometry}

\newtheorem{thm}{Theorem}

\newcommand{\blind}{1}

\begin{document}

\def\spacingset#1{\renewcommand{\baselinestretch}%
{#1}\small\normalsize} \spacingset{1}

%%%%%%%%%%%%%%%%%%%%%%%%%%%%%%%%%%%%%%%%%%%%%%%%%%%%%%%%%%%%%%%%%%%%%%%%%%%%%%

\if1\blind
{
  \title{\bf On a spiked model for large volatility matrix estimation from noisy high-frequency data}
  \author{Keren Shen\\
  	Department of Statistics and Actuarial Science\\
  	The University of Hong Kong\\
  	Jianfeng Yao \\
  	Department of Statistics and Actuarial Science\\
  	The University of Hong Kong\\
    and \\
	Wai Keung Li \\
	Department of Statistics and Actuarial Science\\
	The University of Hong Kong\\}
  \maketitle
} \fi

\if0\blind
{
  \bigskip
  \bigskip
  \bigskip
  \begin{center}
    {\LARGE\bf On a spiked model for large volatility matrix estimation from noisy high-frequency data}
\end{center}
  \medskip
} \fi

\bigskip
\begin{abstract}
Recently, inference about high-dimensional integrated covariance matrices (ICVs) based on noisy high-frequency data has emerged as a challenging problem. In the literature, a pre-averaging estimator (PA-RCov) is proposed to deal with the microstructure noise. Using the large-dimensional random matrix theory, it has been established that the eigenvalue distribution of the PA-RCov matrix is intimately linked to that of the ICV through the Mar$\check{c}$enko-Pastur equation. Consequently, the spectrum of the ICV can be inferred from that of the PA-RCov. However, extensive data analyses demonstrate that the spectrum of the PA-RCov is {\em spiked}, that is, a few large eigenvalues (spikes) stay away from the others which form a rather continuous distribution with a density function (bulk). Therefore, any inference on the ICVs must take into account this spiked structure. As a methodological contribution, we propose a spiked model for the ICVs where spikes can be inferred from those of the available PA-RCov matrices. The consistency of the inference procedure is established and is checked by extensive simulation studies. In addition, we apply our method to the real data from the US and Hong Kong markets. It is found that our model clearly outperforms the existing one in predicting the existence of spikes and in mimicking the empirical PA-RCov matrices. \\
\end{abstract}

\noindent%
\vfill

\newpage
\spacingset{1.45} % DON'T change the spacing!
\section{Introduction}
\noindent
Modeling and forecasting covariance matrices of asset returns play a crucial role in many financial applications, such as portfolio allocation and asset pricing. With the availability of intraday financial data, it becomes possible to estimate the so-called {\em integrated covariance matrix} (ICV) of asset returns using the {\em realized covariance matrix} which is directly attainable from the high-frequency data (\citealp{ABDL03}, \citealp{BS04} and \citealp{BHLS11}). More precisely, suppose there are $p$ stocks whose log-price processes are denoted by $\mathbf{X}_t = (X_t^1,\cdots,X_t^p)^T$. A commonly used model for $\mathbf{X}_t$ is the following high-dimensional diffusion process:
\begin{displaymath}
d\mathbf{X}_t = \pmb{\mu}_tdt + \mathbf{\Theta}_td\mathbf{W}_t, \quad t \in [0,1],
\end{displaymath}
where $\pmb{\mu}_t = (\mu_t^1,\cdots,\mu_t^p)^T$ is a $p$-dimensional drift process, $\mathbf{\Theta}_t$ is a $p\times p$ co-volatility matrix, and $(\mathbf{W}_t)$ is a $p$-dimensional standard Brownian motion. The ICV is defined as
\begin{displaymath}
\text{ICV} := \int_{0}^{1} \mathbf{\Theta}_t\mathbf{\Theta}_t^T dt.
\end{displaymath}
In practice, the intraday prices are always contaminated by the market microstructure noise. As a matter of fact, we observe, instead of the underlying process $(\mathbf{X}_{t_i})$, a noisy version:
\begin{displaymath}
\mathbf{Y}_{t_i} = \mathbf{X}_{t_i} + \pmb{\varepsilon}_i, \quad i = 1, \cdots,n,
\end{displaymath}
where $\mathbf{Y}_{t_i} = (Y_i^1,\cdots,Y_i^p)^T$ denotes the observations, and $\pmb{\varepsilon}_{i} = (\varepsilon_i^1,\cdots,\varepsilon_i^p)^T$ denotes the noise, which is i.i.d. with mean $\mathbf{0}$ and covariance matrix $\mathbf{\Sigma}_e$, and is independent of $(\mathbf{X}_{t_i})$.\\

\noindent
To filter out the effect of the noise, it is recommended in the literature to pre-average the data as follows (\citealp{CKP10}). Choose a window length $k_n = \lfloor \theta \sqrt{n} \rfloor$ where $\theta \in (0,\infty)$ and a function $g(x) = \text{min}(x,1-x)$ on $[0,1]$. For any process $\mathbf{V} = (\mathbf{V}_t)_{t \ge 0}$, consider the increments (return)
\begin{displaymath}
\Delta_i^n \mathbf{V} = \mathbf{V}_{i/n} - \mathbf{V}_{(i-1)/n}, \text{ for } i =1,\cdots,n.
\end{displaymath}
The pre-averaged process is
\begin{displaymath} 
\bar{\mathbf{V}}_i^n=\sum_{j=1}^{k_n-1}g(\frac{j}{k_n})\Delta_{i+j}^n \mathbf{V}, \text{ for } i=0,\cdots,n-k_n+1.
\end{displaymath}
Therefore, the observed return based on the pre-averaging price is 
\begin{displaymath}
\bar{\mathbf{Y}}_{i}^n = \bar{\mathbf{X}}_{i}^n + \bar{\pmb{\varepsilon}}_{i}^n.
\end{displaymath}
Then, the {\em pre-averaging realized covariance matrix} (PA-RCov) is defined as
\begin{displaymath}
\text{PA-RCov} := \frac{n}{n-k_n+2}\frac{1}{\psi_2k_n}\sum_{i=0}^{n-k_n+1} (\bar{\mathbf{Y}}_{i}^n)(\bar{\mathbf{Y}}_{i}^n)^T,
\end{displaymath}
where $n/(n-k_n+2)$ is a finite sample correction and $\psi_2$ is a correction function related to function $g$. In low dimensional case when the number of the stocks $p$ is small compared to the sample size $n$, it can be proved that the PA-RCov is a consistent estimator of the ICV under an appropriate choice of the window length $k$. However, when the number $p$ becomes large, the PA-RCov is no longer a good estimator even in the simplest case when the co-volatility process remains unchanged. In particular, the spectrum of the PA-RCov will deviate from that of the ICV.\\

\noindent
A few papers have appeared recently in the literature to deal with this challenging problem of estimation of high-dimensional ICV matrices. \citet{WZ10} use thresholding and banding regularization techniques to estimate the ICV under sparsity or decay assumptions on the ICV. \citet{TWYZ11} use a similar approach to construct the realized covariance matrix and incorporate low-frequency dynamics in their modeling. However, few work exists to deal with non-sparse ICV matrices while the microstructure noise is present. In a recent paper, \citet{XZ15} propose a new pre-averaging estimator for the spectrum of such {\em non-sparse} ICV matrices. To make the statistical inference of ICV possible, $(\mathbf{X}_t)$ is required to belong to Class $\mathcal{C}$ such that $\mathbf{\Theta}_t = \gamma_t\mathbf{\Lambda}$ for some constant matrix $\mathbf{\Lambda}$ and $(\gamma_t)$ is a c$\grave{a}$dl$\grave{a}$g function from $[0,1]$ to $\mathbb{R}$. Then, the ICV can be written as
\begin{displaymath}
\text{ICV} = \int_{0}^{1} \gamma_t^2 dt \cdot \breve{\mathbf{\Sigma}}, \quad \text{ where } \breve{\mathbf{\Sigma}} = \mathbf{\Lambda}\mathbf{\Lambda}^T.
\end{displaymath}
In other words, $(\gamma_t)$ serves as a time-varying scaling of the base ICV matrix $\breve{\mathbf{\Sigma}}$. Notice that here the dynamic of the volatility process $\mathbf{\Theta}_t$ depends on one scalar function $\gamma_t$ only. However, a fully general model $\mathbf{\Theta}_t = (\theta_{ij}(t))$ using $p^2$ functions would be impossible to identify. Furthermore, we may later extend such setting to a greater freedom as follows. Stocks are naturally grouped to, say $s$ sectors($s \ll p$). With $p_k$ stocks in $k$th sector($p_1+\cdots +p_s=p$), we may allow a different base volatility matrix $\mathbf{\Lambda}_k(p_k \times p_k)$ for each sector $k(1 \le k \le s)$ associated to a specific dynamic function $\gamma_{kt}$. The overall volatility matrix becomes $\mathbf{\Theta}_t = \text{diag}(\gamma_{kt}\mathbf{\Lambda}_k)_{1 \le k \le s}$ and the corresponding integrated covariance matrix becomes $\text{ICV}=\text{diag}(\int_{0}^{1}\gamma_{kt}^2dt \cdot  \mathbf{\Lambda}_k\mathbf{\Lambda}_k^T)_{1 \le k \le s}$. Nevertheless, in this paper, we focus on the simplest setting and extension to the block-diagonal case above is not difficult. To construct consistent estimator for the above ICV, \citet{XZ15} consider a slightly different scheme of the pre-averaging estimator. Choose a window length $k$ and group the intervals $[(i-1)/n,i/n]$, $i = 1, \cdots, 2k\cdot\lfloor n/(2k)\rfloor$ into $m=\lfloor n/(2k)\rfloor$ pairs of non-overlapping windows. For any process $\mathbf{V} = (\mathbf{V}_t)_{t \ge 0}$, let
\begin{displaymath}
\Delta \mathbf{V}_i = \mathbf{V}_{i/n} - \mathbf{V}_{(i-1)/n}, \bar{\mathbf{V}}_i=\frac{1}{k}\sum_{j=0}^{k-1}\mathbf{V}_{((i-1)k+j)/n}, \text{ and } \Delta \bar{\mathbf{V}}_{2i} = \bar{\mathbf{V}}_{2i} - \bar{\mathbf{V}}_{2i-1}.
\end{displaymath}
Then, the following PA-RCov matrix is considered:
\begin{equation}
\mathcal{B}_m := 3\frac{\sum_{i=1}^{m}|\Delta\bar{\mathbf{Y}}_{2i}|^2}{m}\sum_{i=1}^{m}\frac{\Delta\bar{\mathbf{Y}}_{2i}(\Delta\bar{\mathbf{Y}}_{2i})^T}{|\Delta\bar{\mathbf{Y}}_{2i}|^2},
\end{equation}
where for any vector $\mathbf{v}$, $|\mathbf{v}|$ is its Euclidean norm. In this paper, we will use the version of \citet{XZ15} for the PA-RCov $\mathcal{B}_m$. Recall that the {\em empirical spectral distribution} (ESD) of a symmetric matrix $\mathbf{A}$ of size $p \times p$ is the probability distribution 
\begin{displaymath}
F^{\mathbf{A}} = \frac{1}{p} \sum_{j=1}^{p} \delta_{\lambda_j},
\end{displaymath}
where $(\lambda_j)_{1 \le j \le p}$ are the eigenvalues of $\mathbf{A}$, and $\delta_{x}$ is the Dirac mass. A key ingredient here is that the ESDs of  $\mathcal{B}_m$ and the ICV are related in the high-dimensional case. Precisely, when $p \rightarrow \infty$ and $p/m \rightarrow y \in (0, \infty)$, the ESDs of the ICV and $\mathcal{B}_m$ converge almost surely to probability distributions $F^{\text{ICV}}$ and  $F^{\mathcal{B}_{\infty}}$, respectively. Moreover, the Stieltjes transform of $F^{\mathcal{B}_{\infty}}$, denoted by $s_{\mathcal{B}_{\infty}}(z)$, satisfies the Mar$\check{\text{c}}$enko-Pastur equation associated with $F^{\text{ICV}}$. As a result, the spectrum of the ICV can be inferred from that of $\mathcal{B}_m$ using this equation. For instance, assume the limiting spectral distribution of the ICV is of the form of a weighted sum of point masses
\begin{displaymath}
F^{\text{ICV}} = \sum_{k=1}^{N} w_k \delta_{x_k},
\end{displaymath}
where $\{x_1<x_2<\cdots<x_N\}$ is a grid of pre-determined points, and $w_k$'s are weights to be estimated from the PA-RCov $\mathcal{B}_m$. Precisely, in the Mar$\check{\text{c}}$enko-Pastur equation linking $F^{\text{ICV}}$ and $F^{\mathcal{B}_{\infty}}$, $F^{\mathcal{B}_{\infty}}$ is replaced by $F^{\mathcal{B}_m}$ which is available, and the weights $\{w_k\}$ are selected by solving the equation. Note that as a consequence, this estimator of the ICV provides the estimated weights $\{\hat{w}_k\}$ over the grid points $\{x_1<x_2<\cdots<x_N\}$, and these are not exactly the set of eigenvalues of the ICV. For more information about the Stieltjes transform and the Mar$\check{\text{c}}$enko-Pastur equation, please refer to \citet{BS10}.\\

\begin{figure}[H]
	\centering
	\includegraphics[width=11cm]{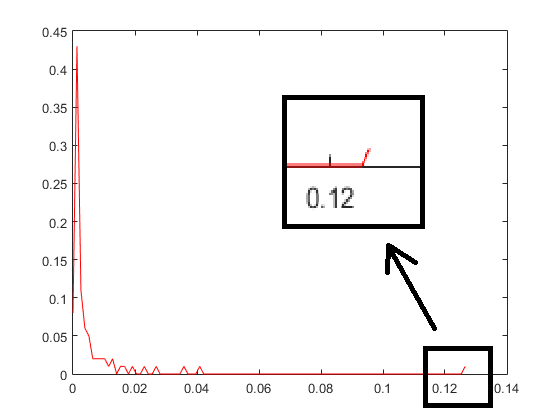}
	\caption{The ESD of the pre-averaging estimator $\mathcal{B}_m$ on September 29th, 2008}
	\label{figureintro}
\end{figure}

\noindent
From the Mar$\check{\text{c}}$enko-Pastur equation, it is known that the eigenvalues of the PA-RCov have a finite support, made of a finite number of intervals. However, \citet{LCPB00} and \citet{PGRAGS02} find that the sample covariance matrix of asset returns always exhibits some large eigenvalues, or {\em spikes}, separated from the core spectrum ({\em bulk}). They also find that the eigenvectors corresponding to these spiked eigenvalues are relatively stable and carry meaningful economical information. For example, the largest eigenvalue corresponds to the largest potential risk of the portfolio. More importantly, our empirical study finds that the spectrum of the PA-RCov also comprises a spike part and a bulk part. For illustration, consider an example of 92 stocks extracted from the S$\&$P 100 index. On September 29th, 2008, the ESD of $\mathcal{B}_m$ made with its 92 eigenvalues is plotted in Figure \ref{figureintro}.

\noindent
From Figure \ref{figureintro}, it is seen that the majority of the 92  eigenvalues concentrate in the range $[0, 0.02]$ while there is a large eigenvalue near $0.13$, far away from the bulk part. Naturally, there should be a corresponding spike for the ICV, which is ought to be tackled separately from the bulk part. Notice that there might be other spikes between the bulk range $[0, 0.02]$ and the largest eigenvalue, though a precise inference on this question is yet to be defined (see Section 4.1.1 for details). Nevertheless, the model proposed by \citet{XZ15} ignores such spiked structure in the spectrum of the ICV as their estimator makes no distinction between the spikes which are large and isolated eigenvalues and the bulk which typically has a continuous density. In this paper, we propose a spiked model for the ICV which is capable of taking into account both the spiked eigenvalues and the bulk component in its spectrum. Such spiked structure in the ICV will lead to an associated spiked structure in the spectrum of the observed PA-RCov $\mathcal{B}_m$ and as demonstrated in Figure \ref{figureintro}, this matches better the reality.\\

\noindent
The spiked model is first introduced by \citet{J01} where all eigenvalues of the population covariance matrix are unit except a fixed small number of them. \citet{BS06} consider a more general spiked model with complex or real and not necessarily Gaussian random variables. Later, \citet{P07} and \citet{BY08} establish the central limit theorem for extreme sample eigenvalues generated by spiked eigenvalues. Moreover, \citet{BY12} consider a {\em generalized spiked eigenvalue model} with a much extended structure of spiked eigenvalues. Other related works include \citet{PY14} and \citet{YBZ15}. The above spiked models all assume that the observed vectors are purely drawn from the multivariate distributions. On the contrary, we construct a spiked model in the setting of \citet{XZ15} where the observed returns are contaminated with finite variation and microstructure noise parts. In this paper, we establish the asymptotic relationship between the spiked eigenvalues of the PA-RCov and those of the ICV, by an equation similar to that in the generalized spiked model introduced by \citet{BY12}. As a result, the spikes of the ICV can be inferred from those of the PA-RCov. The consistency proved in the theorem is also demonstrated by extensive simulation studies. It is found that the asymptotic property holds, in various kinds of situations, where we have one or more spikes, and where the spikes have different magnitudes. In addition, we apply our model to the real data in the US market and the Hong Kong market. We find that our model consistently outperforms the model proposed by \citet{XZ15} in predicting the existence of spikes and in mimicking the empirical PA-RCov matrices.\\

\noindent
In summary, the main contributions of this paper are as follows. To our best knowledge, this is the first paper introducing a spiked model for the realized covariance matrix and the ICV. In addition, we apply the model to analyzing real data, which is not found in \citet{XZ15}. We also report some interesting findings in the empirical study, for example, the magnitude of the largest eigenvalue of the PA-RCov may be a potential leading factor of the volatility of the return series. Finally, we show in detail how and why the model of \citet{XZ15} fails to capture the truth of spikes.\\ 

\noindent
The rest of the paper is organized as follows. Section 2 introduces our new spiked model and establishes the main theorem on the consistency of the estimated spikes. This consistency is checked through extensive simulation studies in Section 3. We apply our model to the real data in Section 4 where we compare our model with that of \citet{XZ15}. Section 5 concludes.\\

\section{A spiked model for the ICV}
\noindent
We set some notations that will be used throughout the paper. For any real matrix $\mathbf{A}$, $\|\mathbf{A}\| = \sqrt{\lambda_{\text{max}}(\mathbf{AA}^T)}$ denotes its spectral norm, where $\lambda_{\text{max}}$ denotes the largest eigenvalue. Moreover, $\mathbf{A}_i$ denotes the $i$th column of $\mathbf{A}$ and $a_i^j$ denotes the $j$th entry in $\mathbf{A}_i$. Finally, $\Gamma_{\mu}$ denotes the support of a finite measure $\mu$ on $\mathbb{R}$.\\ 

\noindent
As said in Introduction, we separate the spectrum of the ICV into a bulk part and a spikes part. Specifically, we assume that the constant matrix component of the ICV,  $\breve{\mathbf{\Sigma}} = \mathbf{\Lambda\Lambda}^T$ has the following form:
\begin{equation}
\breve{\mathbf{\Sigma}} = \begin{pmatrix}
\mathbf{V}_s & \mathbf{0} \\
\mathbf{0} & \mathbf{V}_p  
\end{pmatrix}
,
\end{equation}  
where $\mathbf{V}_s$ is of size $K \times K$, $\mathbf{V}_p$ is of size $(p-K) \times (p-K)$ and $K$ is a fixed number. The eigenvalues of $\mathbf{V}_s$ are $\breve{\alpha}_1 > \cdots > \breve{\alpha}_K>0$ which are the $K$ {\em spikes} of the ICV, and the eigenvalues of $\mathbf{V}_p$ are the {\em bulk} of the ICV. Naturally, it will be assumed that the spikes and the bulk do not overlap with each other. Such spiked spectrum of the ICV will generate a corresponding spiked structure for the PA-RCov $\mathcal{B}_m$, which is much in accordance with the reality.\\

\noindent
The main task here is to infer the spiked eigenvalues of the ICV from those observed in the PA-RCov. This is achieved via the main theorem below which establishes a complex correspondence between the two sets of spikes. This result is established under two sets of assumptions. The first set of assumptions A1 - A8 follow the settings in \citet{XZ15}, that guarantee the consistency of the sequence $\mathcal{B}_m$. They are detailed in the supplementary material (Section B) with the proof of the theorem. The second set of assumptions B1 - B2 below are introduced to tackle the spiked structure of the ICV.
\begin{enumerate}
	\item[B1.] $\breve{\mathbf{\Sigma}} = \mathbf{\Lambda\Lambda}^T$ has the from in (2) where 
	(i) $\mathbf{V}_s$ is of size $K \times K$ where $K$ is a fixed number. The eigenvalues of $\mathbf{V}_s$ are $\breve{\alpha}_1 > \cdots > \breve{\alpha}_K>0$. (ii) As $p \rightarrow \infty$, the ESD $H_p$ of $\mathbf{V}_p$ converges in distribution to a nonrandom limiting distribution $\breve{H}$. (iii) The eigenvalues $\{\beta_{pj}\}$ of $\mathbf{V}_p$ are such that
	\begin{displaymath}
	\text{sup}_jd(\beta_{pj}, \Gamma_{\breve{H}}) = \varepsilon_p \rightarrow 0,
	\end{displaymath}
	where $d(x, A)$ denotes the distance of $x$ to a set $A$ and $\Gamma_{\breve{H}}$ is the support of $\breve{H}$.
	\item[B2.] The sequence $(\breve{\mathbf{\Sigma}})_p$ is bounded in the spectral norm.
\end{enumerate}

\noindent
The main theoretical contribution of the paper is the following theorem which connects the spikes of $\mathcal{B}_m$ to those of the ICV. The proof of the theorem is given in the supplementary material (Section B). \\

\begin{thm}
	Suppose that for all $p$, $(\mathbf{X}_t)$ is a $p$-dimensional process in Class $\mathcal{C}$ for some drift process $\pmb{\mu}_t$ and co-volatility process $(\mathbf{\Theta}_t) = (\gamma_t\mathbf{\Lambda})$. Suppose also that Assumptions A1 - A8 and B1 - B2 hold. Define $\zeta = \int_{0}^{1} (\gamma_t^*)^2dt$, where $\gamma_t^*$ is defined in Assumption A4, and
	\begin{equation}
	\psi(\alpha) = \alpha + y\int \frac{t\alpha}{\alpha-t} dH(t),
	\end{equation}
	where $H(x) = \breve{H}(x/\zeta) \text{ for all }x \ge 0$. Then, for a spiked eigenvalue $\alpha_j$ of the ICV satisfying
	\begin{displaymath}
	\psi'(\alpha_j) > 0,
	\end{displaymath}
	there is an eigenvalue $\lambda_j$ of $\mathcal{B}_m$ such that
	\begin{equation}
	\lambda_j \stackrel{a.s.}{\rightarrow} \psi(\alpha_j).
	\end{equation}
\end{thm}

\bigskip
\noindent
The main setting for Theorem 1 is close to that of \citet{BY12} with however one major difference in the proof: \citet{BY12} assume that samples are drawn from a purely multivariate normal distribution, while the stock returns also contain components from the finite variation process and the microstructure noise. As a result, these two parts should be dealt with before using the standard spiked model theory.\\

\noindent
Therefore, according to Theorem 1, each spike $\alpha_j$ of the ICV will correspond to a spike $\lambda_j$ of the PA-RCov, which tends to $\psi(\alpha_j)$. A natural estimator for $\alpha_j$ would be simply $\tilde{\alpha}_j = \psi^{-1}(\lambda_j)$. However, such direct inversion of the $\psi(\cdot)$ function suffers from many numerical instabilities. Following \citet{BD12}, a more stable estimator can be found using the following relationship
\begin{equation}
\alpha_j \underline{s}(\psi_j) = -1,
\end{equation}
where $\psi_j=\psi(\alpha_j)$ and $\underline{s}(\cdot)$ denotes the companion Stieltjes transform of $F^{\mathcal{B}_{\infty}}$. For general information on Stieltjes transform and large sample covariance matrix, refer to \citet{BS10}. Indeed, Equation (5) is completely equivalent to (3). For details, please refer to \citet{BD12}. The advantage of (5) is that the Stieltjes transform $\underline{s}(\cdot)$ can be directly evaluated using the sample eigenvalues $(\lambda_j)$ as follows. Let $J$ be the set of indexes of $\alpha_j$'s. Then for each $z \notin \Gamma_{F^{\mathcal{B}_{\infty}}}$, we have
\begin{displaymath}
\underline{s}_m^*(z) = -\frac{1-p/m}{z} + \frac{1}{m} \sum_{k \notin J} \frac{1}{\lambda_k - z} \stackrel{a.s.}{\rightarrow} \underline{s}(z).
\end{displaymath}
Therefore, since $\lambda_j$ converging to $\psi(\alpha_j)$ by Theorem 1, define
\begin{equation}
b_{m,j} := -\frac{1-p/m}{\lambda_j} + \frac{1}{m} \sum_{k \notin J} \frac{1}{\lambda_k - \lambda_j}.
\end{equation}
which provides a consistent estimator of $\underline{s}(\psi_j)$. Consequently, we have 
\begin{equation}
-\frac{1}{b_{m,j}} \stackrel{a.s.}{\rightarrow} \alpha_j.
\end{equation}

\noindent
As a result, we define our estimator of the ICV spike $\alpha_j$ to be
\begin{displaymath}
\hat{\alpha}_j = -\frac{1}{b_{m,j}}
\end{displaymath}
where $b_{m,j}$ is defined in (6). Theorem 1, or equivalently Equation (7) thus establishes the strong consistency of $\hat{\alpha}_j$. The proof of Theorem 1 is provided in the supplementary document.\\

\section{Simulation studies}
\begin{figure}[H]
	\centering
	\includegraphics[width=12cm]{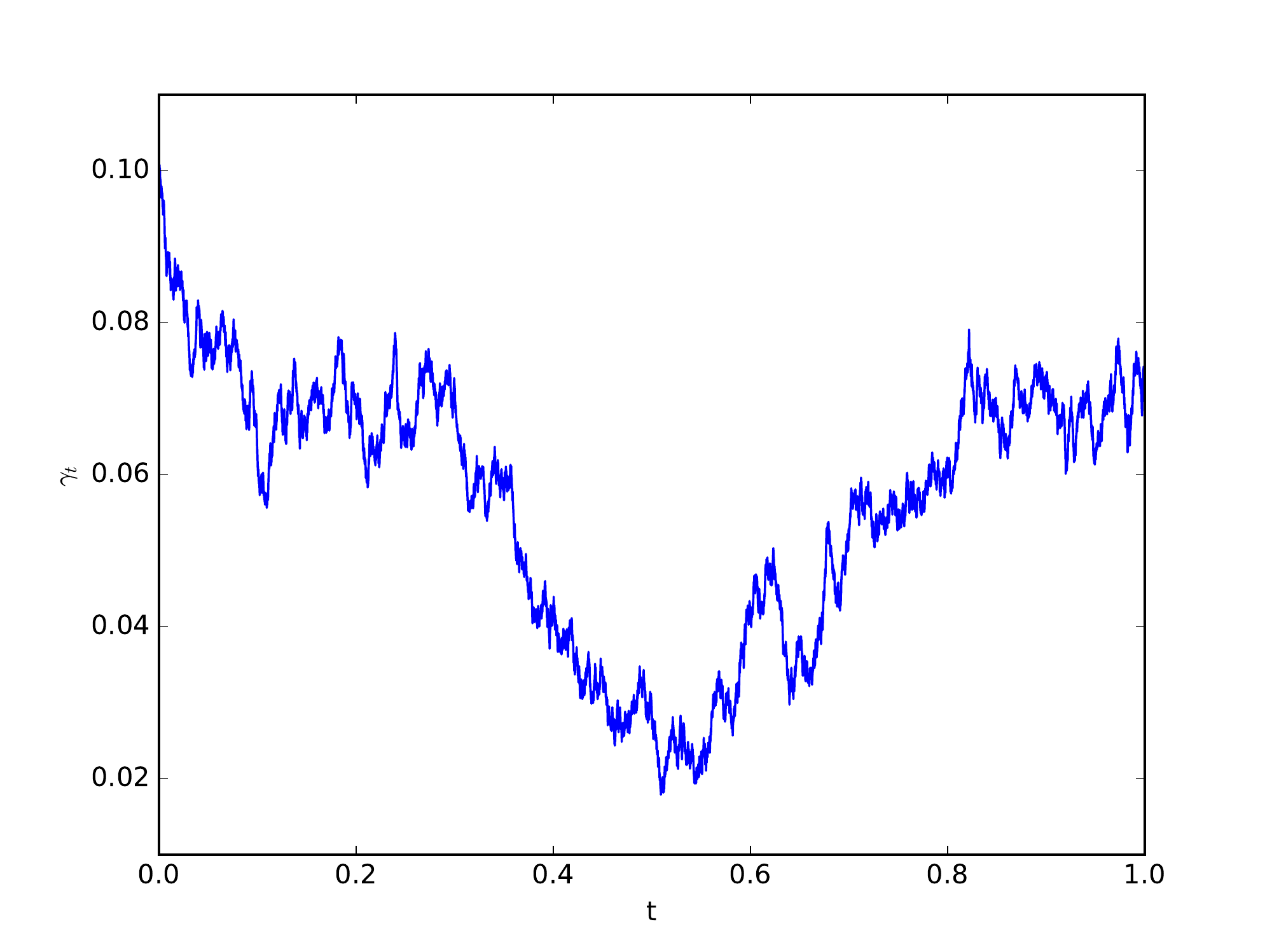}
	\caption{A sample path of the $\gamma_t$ process}
	\label{samplepath}
\end{figure}

\noindent
In this section, we check the finite-sample behavior of Theorem 1 by extensive simulations. We follow the procedure in \citet{XZ15} to construct the simulated log-price process, which satisfies the assumptions we use. In particular, $(\gamma_t)$ follows:
\begin{displaymath}
d\gamma_t = -\rho(\gamma_t-\mu_t)dt + \sigma d\tilde{W}_t, \quad \text{for } t \in [0,1],
\end{displaymath}
where $\rho=10$, $\sigma=0.05$,
\begin{displaymath}
\mu_t = 2\sqrt{0.0009+0.0008 \text{ cos}(2\pi t)},
\end{displaymath}
and $\tilde{W}_t$ is a standard Brownian motion. A sample path of $(\gamma_t)$ is given in Figure \ref{samplepath}.

\noindent
Additionally, the matrix $\breve{\pmb{\Sigma}}$ is in the form of $\pmb{UDU}^T$ where $\pmb{U}$ is a randomly generated unitary matrix and $\pmb{D}$ is a diagonal matrix. We set the first few diagonal entries as spikes and the rest of them are drawn independently from the Beta(1,3) distribution. For example, if there is one spike with a value of $2$ for $\breve{\pmb{\Sigma}}$, then the first diagonal entry of $\pmb{D}$ is set to be $2$ and the others are filled with independent Beta(1,3) samples. Then, the latent log price process $(\mathbf{X}_t)$ follows
\begin{displaymath}
d\mathbf{X}_t = \gamma_t \mathbf{\Lambda}d\mathbf{W}_t, \text{ where } \mathbf{\Lambda} = \breve{\mathbf{\Sigma}}^{1/2},
\end{displaymath}
and $\mathbf{W}_t$ and $\tilde{W}_t$ are independent. Furthermore, the noise $(\pmb{\varepsilon}_i)$ are drawn independently from $N(0, 0.0002\mathbf{I}_p)$. Under such structure, if we treat the largest eigenvalue of the integrated covariance matrix as the signal, then the mean noise-to-signal ratio, or the magnitude of the variation in the noise relative to the true signal, is $0.19\%$ to $1.85\%$, depending on the choice of the spike. Such noise-to-signal ratio is similar to the reality, see \citet{HL06} and \citet{BR08}. Finally, we set $k=\lfloor 0.1 n^{2/3} \rfloor$. \\

\noindent
To illustrate the consistency, we fix the ratio $y=p/m$ to be $1$, or $2$. For each ratio, we set the asset dimension $p$ to be $50$, $100$, $150$, $200$, $250$ and $300$, and use the equations $k=\lfloor 0.1 n^{2/3} \rfloor$ and $m = \lfloor n/(2k) \rfloor$ to solve for the corresponding $n$. Then, the above simulation procedure is conducted and we get the pre-averaging estimator $\mathcal{B}_m$ by Equation (1). Using Equations (6) and (7), we get the estimator for the spikes $(\alpha_j)$ of the ICV. The true value for the spike of the ICV can be attained as follows. We use numerical method to estimate the integral $\int_{0}^{1} \gamma_t^2dt$ which is multiplied to the leading diagonal entries of the matrix $\pmb{D}$ to produce a proxy of the true value for the spikes of the ICV. The whole procedure is repeated $500$ times and the result is shown in Tables 1 to 4.\\

\noindent
We consider four cases: one spike of magnitude $2$, one spike of magnitude $5$, two spikes of magnitude $3$ and $1.5$, respectively, and two spikes of magnitude $15$ and $5$, respectively. Note that these spikes are for $\breve{\pmb{\Sigma}}$, not for the ICV. For each spike $\alpha_j$, we report the empirical bias $|\hat{\alpha}_j-\alpha_j|$, its relative error $|\hat{\alpha}_j-\alpha_j|/|\alpha_j|$ measured in percentage, and the MSE of $\hat{\alpha}_j$. For the 2-spike case, the result in Tables \ref{table3} and \ref{table4} is separated by a comma where the result of the larger spike is before the comma and that of the smaller one is after the comma. For example, in Table \ref{table3} for two spikes $3$ and $1.5$, when $p=100$ and $p/m=1$, $0.0017(15.0)$ is for the larger spike $3$ and $0.0008(13.4)$ is for the smaller one $1.5$. In addition, the percentage of error is shown inside the parentheses.\\

\begin{table}[H]
	\centering
	\caption{Empirical bias, relative error (in $\%$) and MSE of the spike estimates. Case 1: $K=1$, $\alpha_1=2$.}
	\begin{tabular}{|l|c|c|c|c|}
		\hline
		& \multicolumn{2}{|c|}{$p/m=1$} & \multicolumn{2}{|c|}{$p/m=2$} \\
		\hline
		& bias($\%$) & MSE & bias($\%$) & MSE \\
		\hline
		$p=50$ & 0.0014(18.3) & 0.0010 & 0.0018(23.1) & 0.0012 \\
		$p=100$  & 0.0010(12.7) & 0.0007 & 0.0014(18.2) & 0.0010 \\
		$p=150$  & 0.0008(10.7) & 0.0006 & 0.0012(15.2) & 0.0008 \\
		$p=200$  & 0.0007(9.6) & 0.0006 & 0.0010(12.6) & 0.0007 \\
		$p=250$  & 0.0007(8.8) & 0.0005 & 0.0010(12.4) & 0.0007 \\
		$p=300$  & 0.0006(7.6) & 0.0004 & 0.0009(11.3) & 0.0007 \\
		\hline
	\end{tabular}
	\label{table1}
\end{table}

\begin{table}[H]
	\centering
	\caption{Empirical bias, relative error (in $\%$) and MSE of the spike estimates. Case 2: $K=1$, $\alpha_1=5$.}
	\begin{tabular}{|l|c|c|c|c|}
		\hline
		& \multicolumn{2}{|c|}{$p/m=1$} & \multicolumn{2}{|c|}{$p/m=2$} \\
		\hline
		& bias($\%$) & MSE & bias($\%$) & MSE \\
		\hline
		$p=50$ & 0.0051(26.3) & 0.0024 & 0.0057(29.4) & 0.0031 \\
		$p=100$  & 0.0038(19.8) & 0.0019 & 0.0044(23.0) & 0.0025 \\
		$p=150$  & 0.0034(17.4) & 0.0016 & 0.0038(19.7) & 0.0022 \\
		$p=200$  & 0.0029(15.3) & 0.0015 & 0.0032(16.6) & 0.0018 \\
		$p=250$  & 0.0028(14.7) & 0.0013 & 0.0031(15.9) & 0.0018 \\
		$p=300$  & 0.0025(13.1) & 0.0013 & 0.0028(14.7) & 0.0017 \\
		\hline
	\end{tabular}
	\label{table2}
\end{table}

\begin{table}[H]
	\centering
	\footnotesize
	\caption{Empirical bias, relative error (in $\%$) and MSE of the spike estimates. Case 3: $K=2$, $(\alpha_1,\alpha_2)=(3,1.5)$.}
	\begin{tabular}{|l|c|c|c|c|}
		\hline
		& \multicolumn{2}{|c|}{$p/m=1$} & \multicolumn{2}{|c|}{$p/m=2$} \\
		\hline
		& bias($\%$) & MSE & bias($\%$) & MSE \\
		\hline
		$p=50$ & 0.0022(18.6), 0.0010(16.8) & 0.0013, 0.0007 & 0.0025(21.5), 0.0012(20.4) & 0.0016, 0.0009 \\
		$p=100$  & 0.0017(15.0), 0.0008(13.4) & 0.0011, 0.0006 & 0.0019(16.8), 0.0009(16.3) & 0.0013, 0.0007 \\
		$p=150$  & 0.0015(13.1), 0.0006(10.5) & 0.0010, 0.0004 & 0.0017(14.3), 0.0008(13.7) & 0.0012, 0.0006 \\
		$p=200$  & 0.0013(11.0), 0.0005(9.1) & 0.0008, 0.0004 & 0.0015(13.0), 0.0008(13.0) & 0.0011, 0.0006 \\
		$p=250$  & 0.0012(10.2), 0.0005(8.3) & 0.0007, 0.0004 & 0.0013(11.6), 0.0006(11.0) & 0.0010, 0.0006 \\
		$p=300$  & 0.0011(9.4), 0.0005(7.9) & 0.0007, 0.0003 & 0.0013(10.9), 0.0006(10.3) & 0.0009, 0.0005 \\
		\hline
	\end{tabular}
	\label{table3}
\end{table}

\begin{table}[H]
	\centering
	\footnotesize
	\caption{Empirical bias, relative error (in $\%$) and MSE of the spike estimates. Case 4: $K=2$, $(\alpha_1,\alpha_2)=(15,5)$.}
	\begin{tabular}{|l|c|c|c|c|}
		\hline
		& \multicolumn{2}{|c|}{$p/m=1$} & \multicolumn{2}{|c|}{$p/m=2$} \\
		\hline
		& bias($\%$) & MSE & bias($\%$) & MSE \\
		\hline
		$p=50$ & 0.0176(30.9), 0.0031(16.0) & 0.0079, 0.0022 & 0.0181(31.2), 0.0046(23.7) & 0.0095, 0.0033 \\
		$p=100$  & 0.0167(28.8), 0.0029(14.8) & 0.0056, 0.0018 & 0.0165(28.4), 0.0035(18.2) & 0.0077, 0.0025 \\
		$p=150$  & 0.0155(26.8), 0.0026(13.5) & 0.0048, 0.0016 & 0.0155(26.7), 0.0031(16.1) & 0.0065, 0.0020 \\
		$p=200$  & 0.0137(23.7), 0.0026(13.2) & 0.0046, 0.0014 & 0.0140(24.2), 0.0029(15.2) & 0.0056, 0.0018 \\
		$p=250$  & 0.0132(22.8), 0.0023(12.2) & 0.0037, 0.0014 & 0.0133(22.9), 0.0028(14.7) & 0.0056, 0.0018 \\
		$p=300$  & 0.0125(21.6), 0.0022(11.6) & 0.0035, 0.0012 & 0.0126(21.8), 0.0026(13.7) & 0.0050, 0.0017 \\
		\hline
	\end{tabular}
	\label{table4}
\end{table}

\noindent
From Tables \ref{table1} to \ref{table4}, we see that in general, the consistency holds quite well, that is, when the dimension $p$ gets bigger, both the empirical bias and the MSE decrease in all situations. For instance, in Table \ref{table1} for $p/m=1$, the mean percentage of error is reduced from $18.3\%$ when $p=50$, to only $7.6\%$ for the case $p=300$, achieving a quite satisfactory accuracy. In the meanwhile, this clear downward trend for errors for increasing $p$ holds, in all four cases.\\

\noindent
Generally, the error is smaller when $p/m=1$ than that when $p/m=2$. For example, when there are two spikes with values of $3$ and $1.5$, respectively, and $p=300$, the errors are $9.4\%$ and $7.9\%$, respectively for $p/m=1$; they become $10.9\%$ and $10.3\%$, respectively for $p/m=2$. This phenomenon is quite reasonable and may be due to fact that we have more data points to make inference for the value of spike when $p/m=1$, given the same dimension $p$. However, it is found in general that the MSE for the estimation error is quite large, relative to the bias. This can be explained by the complexity of the simulation procedure used which involves continuous time diffusions and their discretization.\\

\noindent
To conclude, the consistency of Theorem 1 for the spikes estimator is strongly confirmed by the extensive simulation studies, for various choices of spikes and their magnitudes, and different combinations of the pair of dimension and sample size.\\

\section{Real data analysis}
\noindent
In order to compare our model with spikes (ModelSp) with that of  \citet{XZ15} without spikes (ModelWs), we apply the two models to the stocks in the US market and the Hong Kong market. We find that ModelSp not only appears to be a more natural choice from the empirical point of view, but also consistently provides more accurate estimation than ModelWs does. Here we present our main findings on the stocks in the US market. Analysis is also carried out for the stocks from the Hong Kong market which is presented in the Appendix A of the supplementary document.\\

\noindent
Precisely, we analyze $92$ stocks which are the available components of the S$\&$P $100$ index in the year 2008. The constituents of the S$\&$P $100$ index represent almost $45\%$ of the market capitalization of the US equity market and most stocks stand for most established and largest companies in the US, which, in our point of view, are able to represent the US large-capitalization stocks. Among the $92$ companies, 16 are in financial, 16 are in services, 14 are in basic materials, 13 are in healthcare, 12 are in technology, 10 are in industrial goods, 9 are in consumer goods and 2 are in utilities. The detailed list is provided in Section C of the supplementary document. We download the tick-by-tick data from the Wharton TAQ database. The sampling period starts from January 3rd, 2008 and ends at December 31st, 2008, totally $253$ trading days. As usually done, only the transactions between 9:30 and 16:00 are considered. Moreover, the following cleaning procedures are adopted before constructing the pre-averaging estimator. If there are multiple transactions in a single second, the median price for that second is retained. Here, we take the median of the prices for simplicity, and other methods may be used, for instance, randomly chosen price in the second, see \citet{JLK16}. For each second between 9:30 and 16:00, if there is no trade, the price of previous trade is taken as the price at that second. Therefore, we have $23401$ prices for a stock in one day, the times 9:30 and 16:00 being included.\\

\noindent
We then take the logarithm of these prices and construct the pre-averaging estimator $\mathcal{B}_m$ for every trading day, by setting $\theta=0.19$ and $\alpha=2/3$.\\

\subsection{Detection of spikes and economic meaning of corresponding eigenvectors}
\noindent
To start with, we follow the approach in \citet{PY14} to estimate the number of spikes. The general idea is as follows. Consider the differences between consecutive eigenvalues of $\mathcal{B}_m$, $\lambda_1 \ge \lambda_2 \ge \cdots \ge \lambda_p$, as:
\begin{displaymath}
\delta_j := \lambda_j -\lambda_{j+1}, \quad j \ge 1.
\end{displaymath}
Define a threshold $d_m$ to be $Cm^{-2/3}\sqrt{2\text{ log log }m}$ where $C$ is a tuning parameter which is data-driven. For the smallest $j$ where $\delta_j < d_m$, the number of spikes of the ICV is estimated to be $\hat{K}=j-1$. The tuning parameter $C$ is set in the following way. The idea is to use the difference of the two largest eigenvalues of a random Wishart matrix, which corresponds to the case without any spikes. Five hundred independent replications are drawn to get an approximated distribution of the difference between the two largest eigenvalues for a random Wishart matrix. The tuning parameter $C$ is taken to be
\begin{displaymath}
C = s \cdot m^{2/3}/\sqrt{2\text{ log log }m},
\end{displaymath}
where the quantile $s$ is estimated by the average of the 10th and the 11th largest spacings among these $500$ replications. For details, please refer to \citet{PY14}.\\

\noindent
We find that there are several spikes for $\mathcal{B}_m$ almost everyday. The 253 numbers of spikes detected during the year 2008 are tabulated in Table \ref{table5}.
\begin{table}[H]
	\centering
	\caption{Distribution of the number of spikes $\hat{K}$ detected daily during 2008}
	\footnotesize
	\begin{tabular}{|l*{2}{c}|}			
		\hline
		$\hat{K}$ & Count & Percent \\
		\hline
		1 & 38 & $15.02\%$ \\
		2 & 95 & $37.55\%$ \\
		3 & 84 & $33.20\%$ \\
		4 & 26 & $10.28\%$ \\
		5 & 9 & $3.56\%$ \\
		6 & 1 & $0.40\%$ \\
		\hline			
	\end{tabular}
	\label{table5}		
\end{table}
\noindent
We find that there are two or three spikes for most days, totally $70.75\%$ of the whole year. A complete plot of these numbers is shown in Figure \ref{figure4}.\\

\noindent
As an example, consider September 29th when the Dow index declines 777.68 points, the largest point drop in history. The corresponding spectrum of $\mathcal{B}_m$ is given in Figure \ref{figureintro}. It is clearly seen that there is a spiked eigenvalue of $\mathcal{B}_m$ around $0.13$, far larger than the rest of the eigenvalues, which confirms the spiked model we propose. Using the above detection method, we also find that $\hat{K}=1$.\\

\noindent
In addition, by analyzing the eigenvectors corresponding to the largest eigenvalues, we find that for the spiked eigenvalues, the eigenvectors do have some economic meanings. Almost all elements in the eigenvector corresponding to the largest eigenvalue share the same sign, which indicates that this eigenvector represents the market component. Moreover, we find the largest elements in the second eigenvector always represent those companies in the field of finance and basic materials, which indicates that the second eigenvector is a "vector of industry fields". However, eigenvectors corresponding to the "bulk" eigenvalues show no significant patterns and behave as random vectors. These findings confirm the conclusion drawn in \citet{LCPB00} and \citet{PGRAGS02}. To illustrate, we take a look at the largest $10$ elements in the eigenvector corresponding to the largest eigenvalue of $\mathcal{B}_m$ on September 29th:
\begin{table}[H]
	\centering
	\caption{Largest ten components of the first eigenvector on September 29th, 2008}
	\footnotesize
	\begin{tabular}{|lll|}			
		\hline
		Stock & Firm & Industry \\
		\hline
		MS	& Morgan Stanley	& Financial\\
		C	& Citigroup Inc &	Financial\\
		BK & Bank of New York	& Financial\\
		BAC	& Bank of America Corp &	Financial\\
		JPM	& JP Morgan Chase $\&$ Co &	Financial\\
		GS & Goldman Sachs & Financial\\
		AAPL & Apple Inc & Consumer goods\\
		AIG	& American International Group	& Financial\\
		USB	& US Bancorp	& Financial\\
		AMZN	& Amazon.com	& Services\\
		\hline		
	\end{tabular}
	\label{table6}		
\end{table}
\noindent
It is found from Table \ref{table6} that the first eigenvector consists of large companies from different industries, sharing the same sign, which reflects the market component. In the meantime, most companies are from the financial field, indicating also an industry signal. We also take a look at those companies corresponding to the largest values on the second to the fifth eigenvectors.\\

\begin{table}[H]
	\centering
	\caption{Largest ten components of the second to fifth eigenvectors on September 29th, 2008}
	\scriptsize
	\begin{tabular}{|lll|lll|}			
		\hline
		Stock & Firm & Industry & Stock & Firm & Industry \\
		\hline
		\multicolumn{3}{|c|}{Second eigenvector} & \multicolumn{3}{|c|}{Third eigenvector} \\
		\hline
		AIG	& American International Group	& Financial & NOV	& National Oilwell Varco & Basic materials \\
		MA & Mastercard & Financial &  AMZN	& Amazon.com	& Services \\
		OXY & Occidental Petroleum Corp & Basic materials &  APC	& Anadarko Petroleum Corp & Basic materials  \\
		APC	& Anadarko Petroleum Corp & Basic materials & DVN	& Devon Energy & Basic materials\\
		NOV	& National Oilwell Varco & Basic materials & BLK	& BlackRock Idec  & Financial \\
		MET	& Metlife Inc  & Financial &  LOW	& Lowe's  & Services \\
		GS & Goldman Sachs & Financial & QCOM	& Qualcomm Inc & Technology \\
		CAT	& Caterpillar Inc	& Industrial goods & EMR	& Emerson Electric Co &	Financial \\
		APA	& Apache Corp	& Basic materials & HD & Home Depot & Services \\
		GM	& General Motors & Consumer goods & HAL & Halliburton & Basic materials \\
		\hline		
		\multicolumn{3}{|c|}{Fourth eigenvector} & \multicolumn{3}{|c|}{Fifth eigenvector} \\
		\hline
		BK & Bank of New York	& Financial & MS	& Morgan Stanley	& Financial \\
		QCOM	& Qualcomm Inc & Technology &  NOV	& National Oilwell Varco &	Basic materials \\
		TWX & Time Warner Inc & Services & AMZN	& Amazon.com	& Services \\
		EMC	& EMC Corp & Technology &  BLK	& BlackRock Idec  & Financial \\
		ALL & Allstate Corp & Financial &  APC	& Anadarko Petroleum Corp & Basic materials \\
		EXC	& Exelon  & Utilities & GS & Goldman Sachs & Financial \\
		CMCSA	& Comcast Corp	& Services & APA & Apache Corp & Basic materials \\
		GILD & Gilead Sciences & Healthcare & HAL & Halliburton & Basic materials \\
		F & Ford Motor & Consumer goods & DVN	& Devon Energy & Basic materials \\
		T	& AT$\%$T Inc & Technology & COP & ConocoPhillips & Basic materials \\
		\hline		
	\end{tabular}
	\label{table7}		
\end{table}
\noindent
No particular pattern is found in Table \ref{table7} and we think it is because these eigenvectors belong to the bulk eigenvalues.

\subsection{The underlying relationship between the magnitude of spikes and the volatility of the S$\&$P 100}
\begin{figure}[H]
	\centering
	\includegraphics[width=9cm]{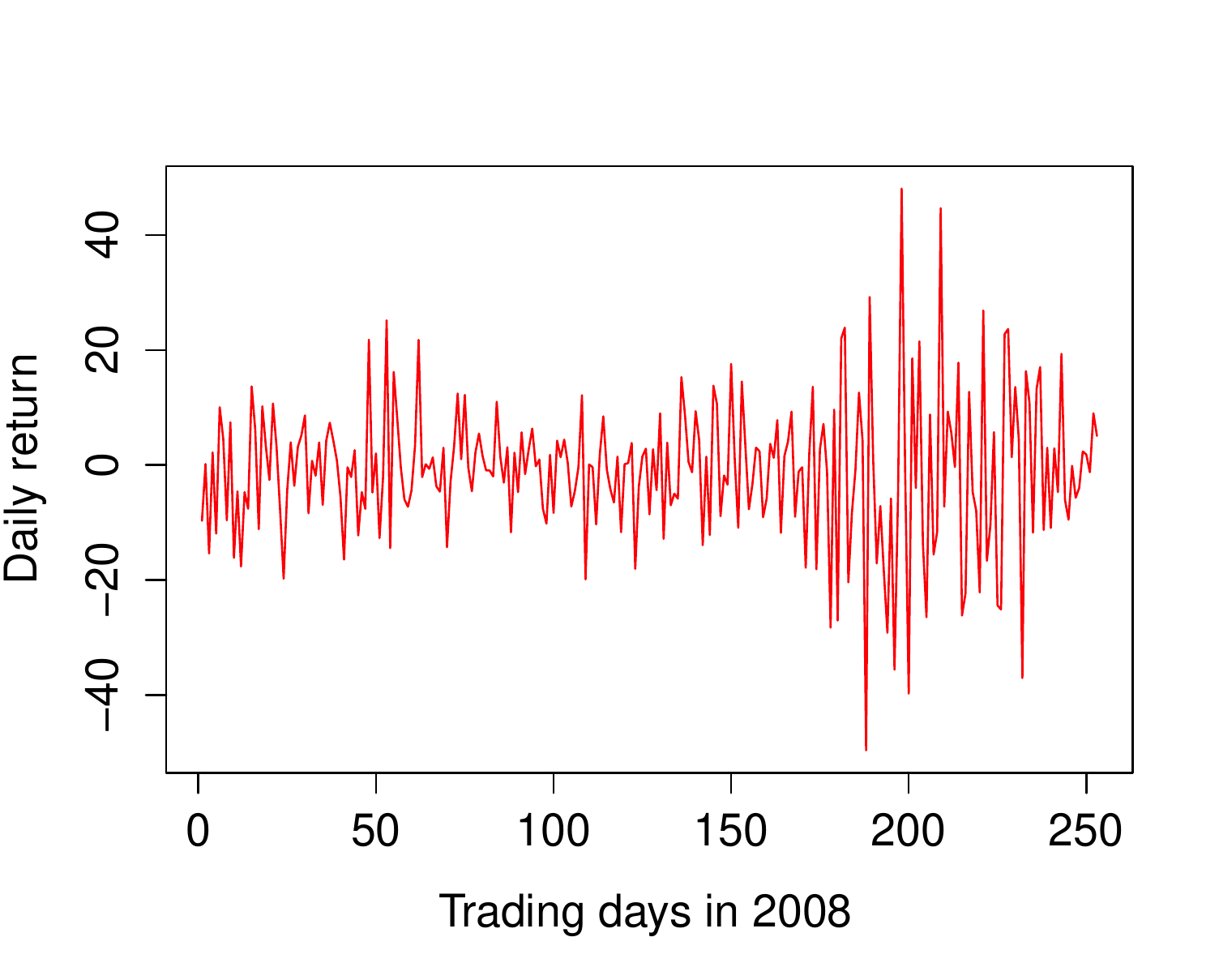}
	\caption{The daily return of the S$\&$P 100 index during the year 2008}
	\label{figure2}
\end{figure}

\begin{figure}[H]
	\centering
	\includegraphics[width=9cm]{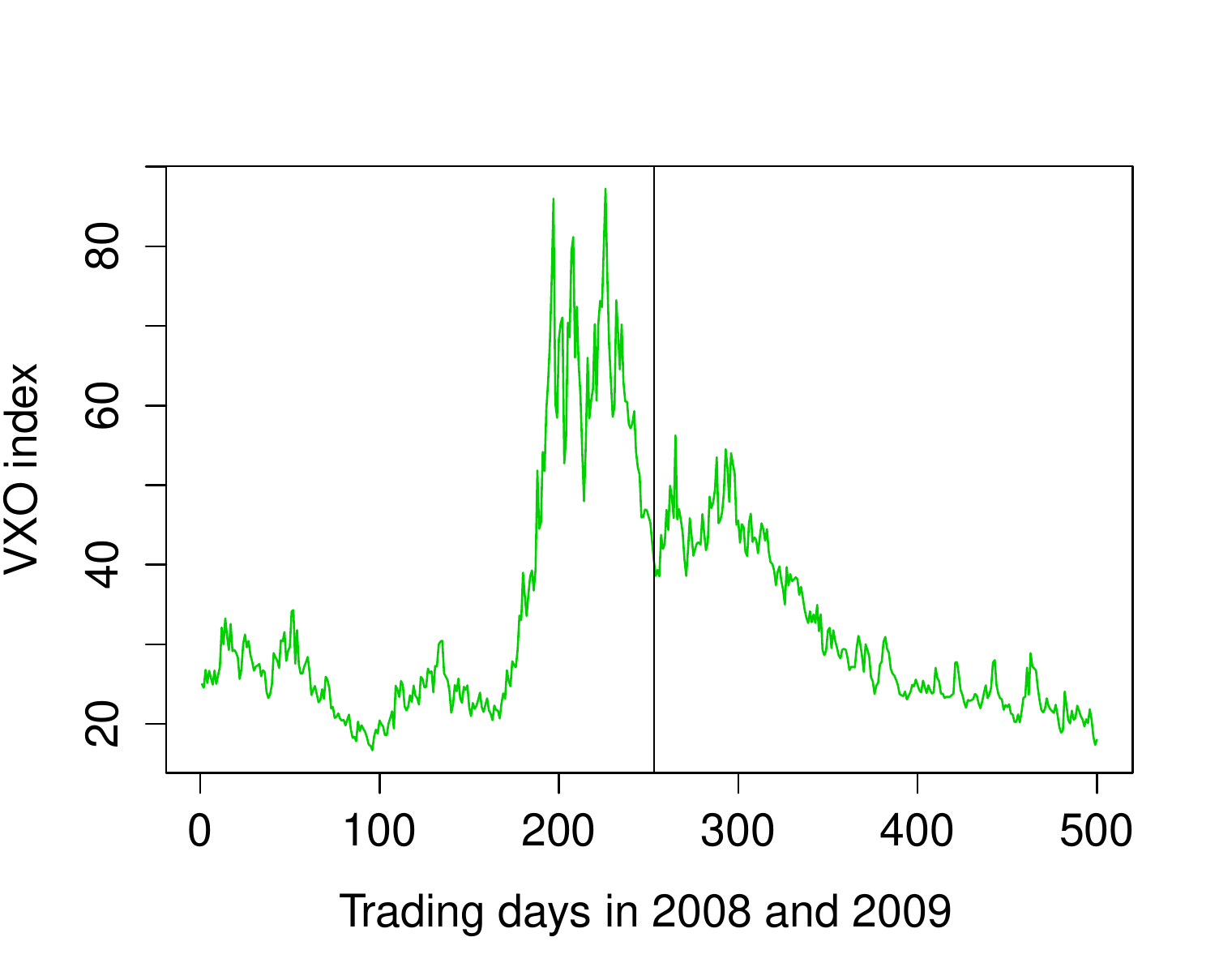}
	\caption{The VXO index during the years 2008 and 2009}
	\label{figure3}
\end{figure}

\noindent
Next, we consider the relationship between the spikes of $\mathcal{B}_m$ and the volatility of the S$\&$P 100 index. We first plot the daily return of the S$\&$P 100 index in the year 2008; and the volatility of the S$\&$P 100 index (VXO) during the years 2008 and 2009, separated by the vertical line in the graph, in Figures \ref{figure2} and \ref{figure3}, respectively.\\

\noindent
From Figures \ref{figure2} and \ref{figure3}, it is seen that during the first eight months (about 170 days) in the year 2008, the return and the volatility curves are relatively in peace. For example, the volatility fluctuates in the interval $[20, 30]$ during the period. However, starting from September 2008, the return series becomes volatile and the volatility reaches the peak during October to November, 2008. The VXO index reaches 80 during this period. After that, the volatility of the S$\&$P 100 index starts going downward and this trend continues into the year 2009. We may also see the downtrend from the graph of the VXO index to the right of the middle line.\\

\begin{figure}[H]
	\centering
	\includegraphics[width=9cm]{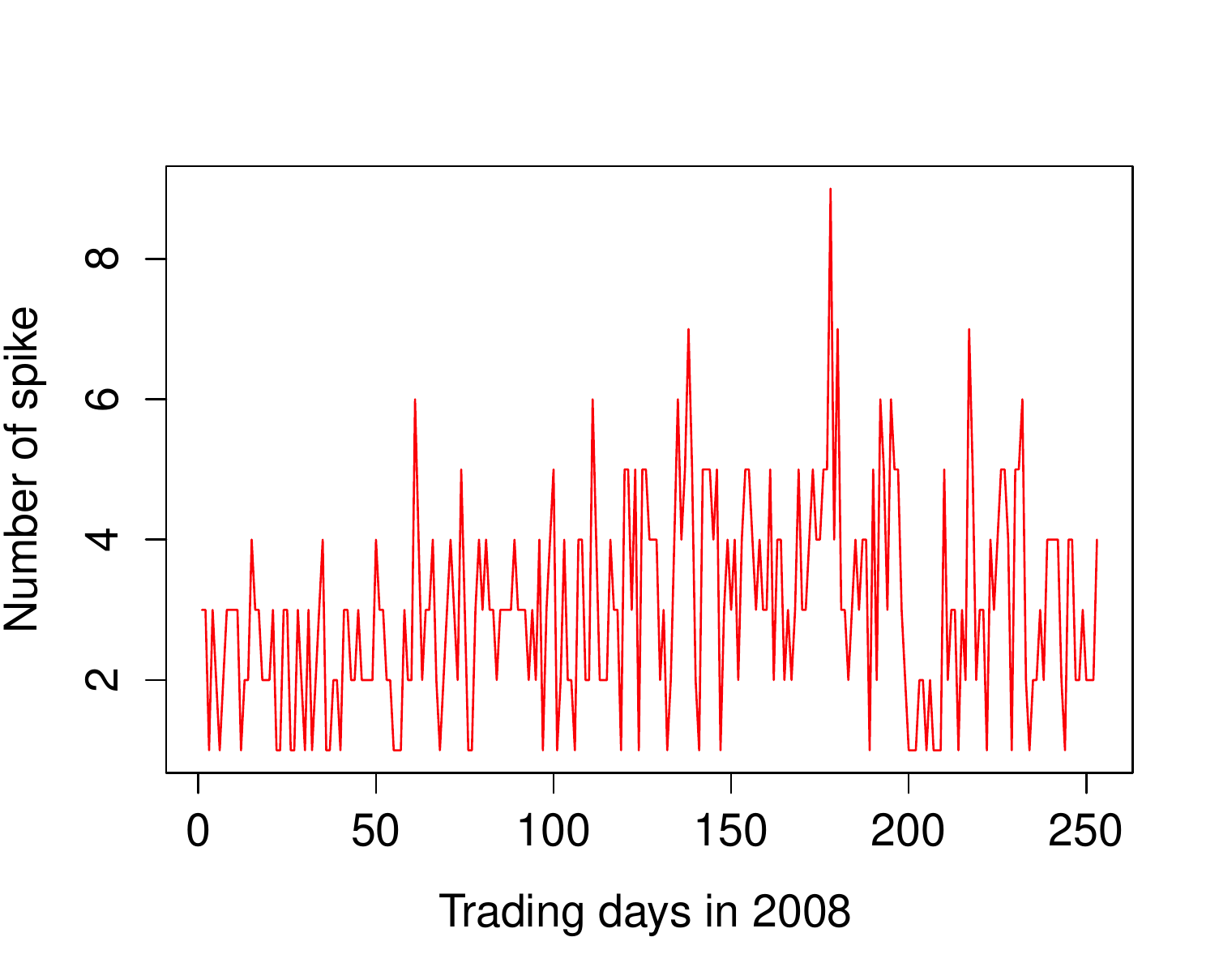}
	\caption{The number of spikes for $\mathcal{B}_m$ during the year 2008}
	\label{figure4}
\end{figure}

\begin{figure}[H]
	\centering
	\includegraphics[width=9cm]{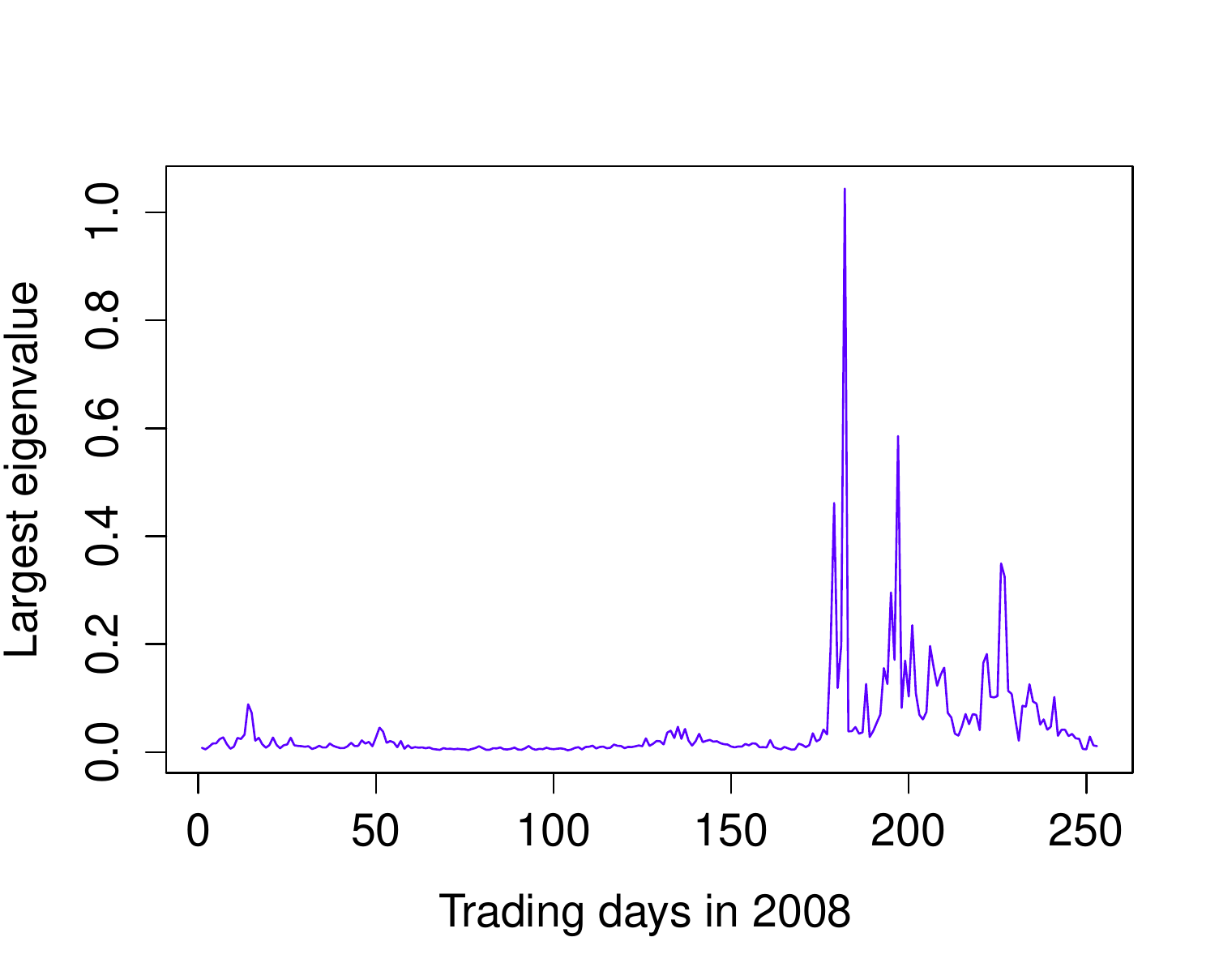}
	\caption{The largest eigenvalue of $\mathcal{B}_m$ during the year 2008}
	\label{figure5}
\end{figure}

\noindent
In addition, we plot the number of spikes for $\mathcal{B}_m$, and the magnitude of the largest eigenvalues for $\mathcal{B}_m$ during the year 2008, in Figures \ref{figure4} and \ref{figure5}, respectively.

\noindent
From Figure \ref{figure5}, we see that the oscillation starts from August, right before the peak of the volatility of the S$\&$P 100 index. Moreover, the relatively small values for the largest eigenvalue before and after the oscillation also appear before the stable situation for the volatility, that is, the small eigenvalues during January to July correspond to the small volatility during February to August; and small eigenvalues after November correspond to the volatility curve which gets smaller gradually starting from December, 2008. In addition, Figures \ref{figure4} and \ref{figure5} suggest that the peak for number of the spikes in the second part of the year seem coincident with that of the magnitude of eigenvalues. As a result, we conjecture that the magnitude of the largest eigenvalue may be a {\em leading indicator} of the volatility.\\

\begin{figure}[H]
	\centering
	\includegraphics[width=9cm]{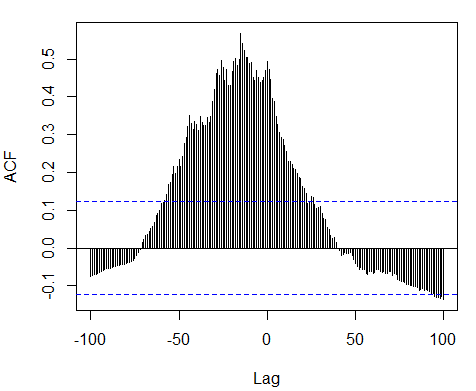}
	\caption{The cross correlation between the magnitude of the largest eigenvalue and the VXO index}
	\label{figure6}
\end{figure}

\noindent
To confirm the above speculation, we calculate the cross correlation between the magnitude of the largest eigenvalue and the VXO index, which is shown in Figure \ref{figure6}. It is found that the most dominant cross correlations occur at lag $-15$, at about $0.57$. In other words, the magnitude of the largest eigenvalue leads the VXO index for 15 trading days, or about three weeks. This is in agreement with the conjecture before, that the oscillation of the largest eigenvalue appears at the 182th trading day, about three weeks before the peak of the VXO at the 197th trading day.

\noindent
In conclusion, our empirical findings described here suggest that the magnitude of the spikes can be treated as a leading indicator of the volatility of the S$\&$P 100 index. It makes sense since the largest eigenvalue always corresponds to the largest possible risk of the portfolio, which certainly reflects the possible volatility of the portfolio of the S$\&$P 100 index.

\subsection{Comparison of models with or without spikes}
\begin{figure}[H]
	\centering
	\begin{subfigure}[b]{0.4\textwidth}
		\includegraphics[width=\textwidth]{rcoveig_9_29.png}
		\caption{Full version}
	\end{subfigure}
	\begin{subfigure}[b]{0.4\textwidth}
		\includegraphics[width=\textwidth]{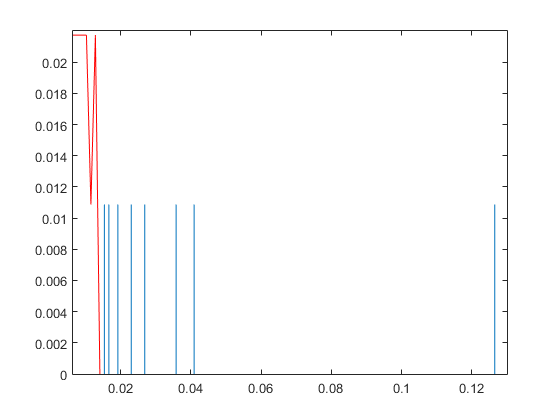}
		\caption{Detailed version}
	\end{subfigure}
	\caption{The ESD of $\mathcal{B}_m$ on September 29th, 2008}
	\label{figure7}
\end{figure}

\noindent
To compare ModelSp and ModelWs, we take two approaches, from empirical and statistical points of view, respectively. For every trading day, we use Equations (6) and (7) to estimate the spikes for ModelSp, and use the method in \citet{XZ15} to estimate the bulk part of eigenvalues for both ModelSp and ModelWs. As said in the Introduction, their approach leads to a family of weights $(w_1, w_2, \cdots, w_N)$ on a grid of points $\{x_1 < x_2 < \cdots < x_N \}$.
\begin{enumerate}
	\item [(a)] Qualitative comparison
\end{enumerate}
\noindent
To construct the ESD of the ICV using ModelSp, we first transform the spikes we detect into weights by assigning a weight of $1/p$ to each spike. Then, assuming there are $M$ spikes detected, the largest $k$ is found such that $\sum_{i=k}^{N} w_i > M/p$, and we set $w_k = \sum_{i=k}^{N} w_i - M/p$ and set $w_j = 0$, $j = k+1,\cdots,N$. Finally, we find the nearest grid point $x_i$ for each spike and set the weight $w_i$ to $1/p$.  For each ESD, we plot a full version and magnified one for the spikes. For illustration, we randomly take September 29th and April 29th as examples, where one and three spikes are detected, respectively.
\begin{enumerate}
	\item [1.] September 29th
\end{enumerate}
We firstly take a look at the ESD of $\mathcal{B}_m$ in Figure \ref{figure7}. It is seen from the full version that most of the weights amass in the small range from $0$ to $0.02$. In addition, there is one large spike out of the bulk part, around $0.125$.

\begin{figure}[H]
	\centering
	\begin{subfigure}[b]{0.4\textwidth}
		\includegraphics[width=\textwidth]{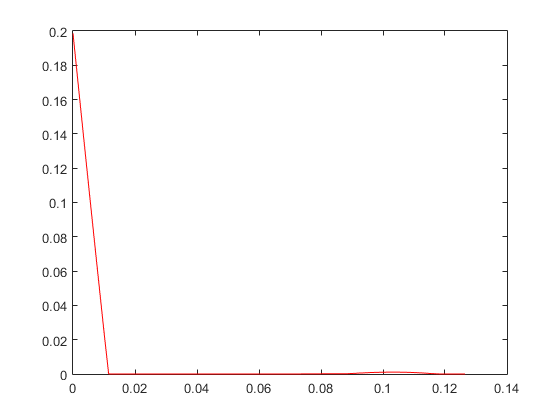}
		\caption{Full version}
	\end{subfigure}
	\begin{subfigure}[b]{0.4\textwidth}
		\includegraphics[width=\textwidth]{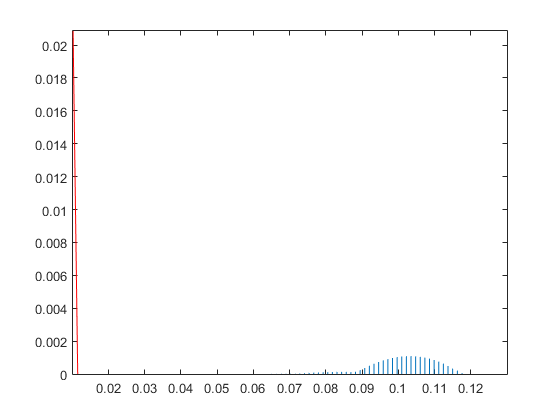}
		\caption{Detailed version}
	\end{subfigure}
	\begin{subfigure}[b]{0.4\textwidth}
		\includegraphics[width=\textwidth]{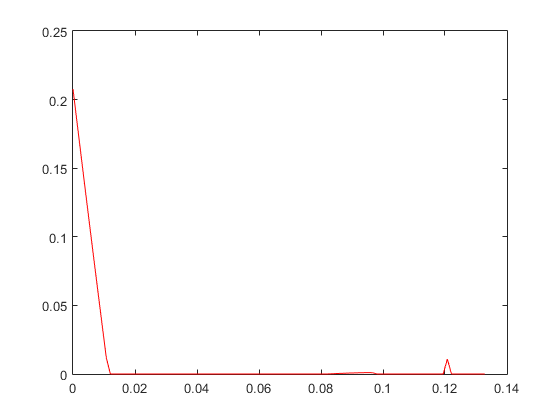}
		\caption{Full version}
	\end{subfigure}
	\begin{subfigure}[b]{0.4\textwidth}
		\includegraphics[width=\textwidth]{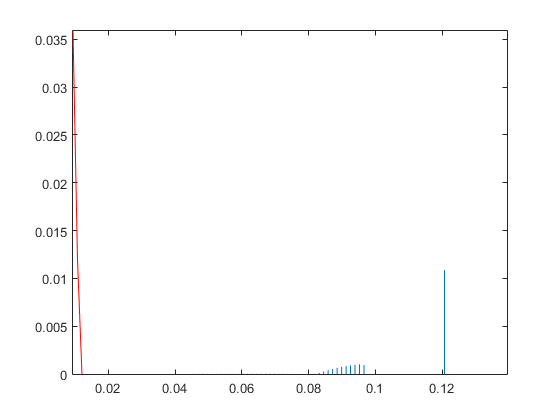}
		\caption{Detailed version}
	\end{subfigure}
	\caption{The ESD of the ICV estimated by ModelWs (top row) and by ModelSp (bottom row) on September 29th, 2008}
	\label{figure8}
\end{figure}

\noindent
For the ESD of the estimated ICV, we firstly look at top row in Figure \ref{figure8} generated by ModelWs. From Figure \ref{figure8}(a), we see that over $95\%$ of weights concentrate in the small values ranging from $0$ to $0.01$. We also witness some really small weights around $0.1$, compared with the bulk part. From Figure \ref{figure8}(b), it is seen that those grouped small weights indicate a continuous component, apart from the bulk.\\

\begin{figure}[H]
	\centering
	\begin{subfigure}[b]{0.4\textwidth}
		\includegraphics[width=\textwidth]{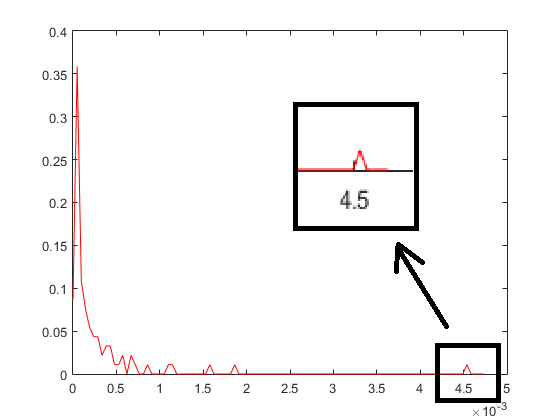}
		\caption{Full version}
	\end{subfigure}
	\begin{subfigure}[b]{0.4\textwidth}
		\includegraphics[width=\textwidth]{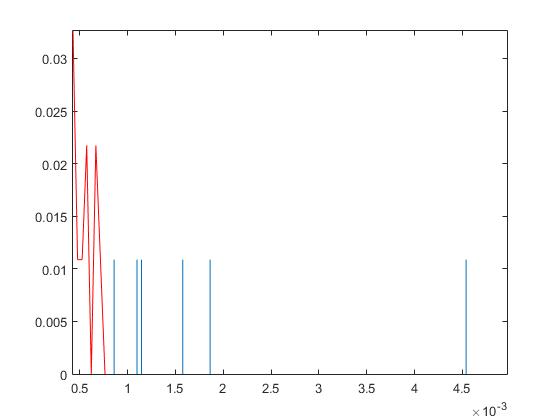}
		\caption{Detailed version}
	\end{subfigure}
	\caption{The ESD of $\mathcal{B}_m$ on April 29th, 2008}
	\label{figure10}
\end{figure}

\begin{figure}[H]
	\centering
	\begin{subfigure}[b]{0.4\textwidth}
		\includegraphics[width=\textwidth]{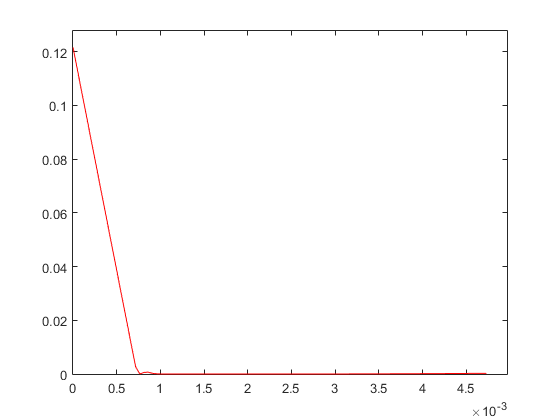}
		\caption{Full version}
	\end{subfigure}
	\begin{subfigure}[b]{0.4\textwidth}
		\includegraphics[width=\textwidth]{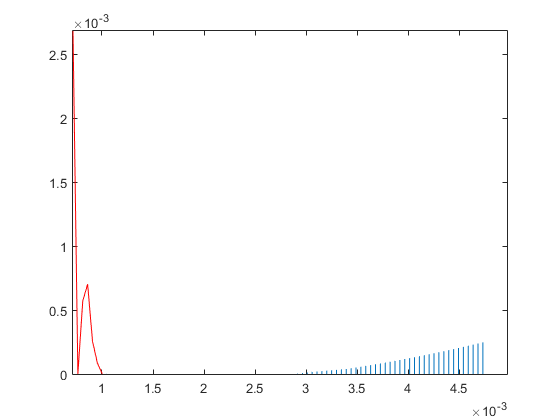}
		\caption{Detailed version}
	\end{subfigure}
	\begin{subfigure}[b]{0.4\textwidth}
		\includegraphics[width=\textwidth]{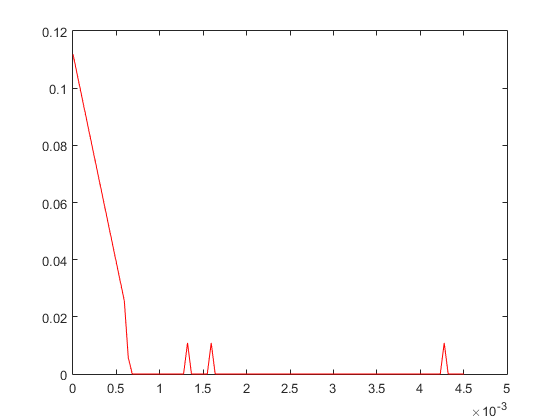}
		\caption{Full version}
	\end{subfigure}
	\begin{subfigure}[b]{0.4\textwidth}
		\includegraphics[width=\textwidth]{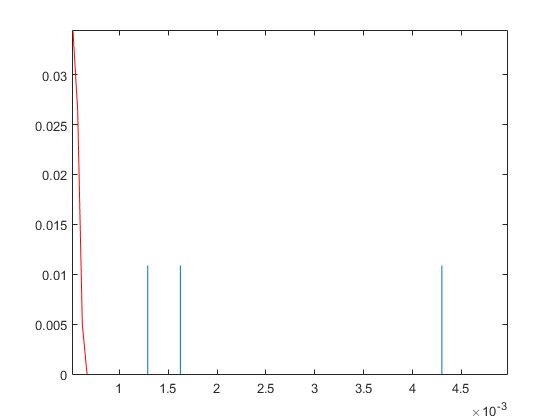}
		\caption{Detailed version}
	\end{subfigure}
	\caption{The ESD of the ICV estimated by ModelWs (top row) and by ModelSp (bottom row) on April 29th, 2008}
	\label{figure11}
\end{figure}

\noindent
For the ESD of the estimated ICV of ModelSp, the bottom row presents a totally different picture from that of ModelWs. There is one spike outside the bulk part and there are some negligibly small weights, compared with the spike. The "bulk + spikes" structure is clearly demonstrated here and is consistent with the ESD of $\mathcal{B}_m$, which is not shared by the ESD of the ICV generated by ModelWs.
\begin{enumerate}
	\item [2.] April 29th
\end{enumerate}
\noindent
Next, we analyze the data on April 29th in Figures \ref{figure10} and \ref{figure11}. We see that ModelSp catches the three spikes in the ESD of $\mathcal{B}_m$. In contrast, ModelWs produces a small smooth density in the range from $0.003$ to $0.005$, which we consider as a counterpart of the spike $0.0045$ of $\mathcal{B}_m$. However, ModelWs is not able to detect other spikes, giving zero weight instead.

\noindent
In conclusion, ModelWs is able to detect the largest spike, which is estimated by a locally smooth continuous density around that spike. However, ModelWs frequently ignores weaker spikes, which are closer to the bulk part. On the other hand, ModelSp successfully detects all the spikes empirically seen in the spectrum of $\mathcal{B}_m$ so that the estimated spectrum by ModelSp clearly shows the "bulk + spikes" structure, which is consistent with that of $\mathcal{B}_m$.
\begin{enumerate}
	\item [(b)] Quantitative comparison
\end{enumerate}
\noindent
In addition, we compare the two models from the statistical point of view. The idea is similar to a parametric bootstrap. Given the two ICV estimates by ModelWs and ModelSp, respectively, we generate by simulation the stock price processes, and obtain the corresponding PA-RCov matrices $\mathcal{B}_{m,Sp}$ and $\mathcal{B}_{m,Ws}$, for the two models, respectively. These simulated matrices are then compared to the observed PA-RCov $\mathcal{B}_m$. The smaller the difference is, the better the model will be. Precisely, after getting the weight $(w_1, w_2, \cdots, w_N)$,  we transform the weight on the grid into a set of estimated eigenvalues by the weight accumulation technique. We start with the smallest value of those points, $x_1$. We set $\lfloor w_1 \times p \rfloor$ eigenvalues to be $x_1$. Then, for the last $x_i$ that we set to be eigenvalue, we find the smallest $x_j$ such that $\lfloor (w_{i+1}+\cdots+w_j) \times p \rfloor > 0$, and set such number of eigenvalues to be $x_j$. The above procedure produces $p$ eigenvalues for ModelWs given the weight $(w_1, w_2, \cdots, w_N)$. For ModelSp, we firstly do exactly the same thing as above to get $p$ eigenvalues, and then replace the largest eigenvalues with those estimated from Equations (6) and (7), which also produces $p$ eigenvalues. Note that only the largest few eigenvalues are different for the two models. We find that for the one spike case, there is only a slight difference between the two models. Moreover, for the case where there are two or more spikes, there is no big difference between the two models regarding the largest eigenvalue, which is below $10\%$. However, for other spikes, the difference between the two models is large, which is above $25\%$.\\

\noindent
Since we do not know the true eigenvalues of the ICV, we compare the estimated ones as follows. As in Section 3, we generate the path for $(\gamma_t)$ and evaluate the integral $\int_{0}^{1} \gamma_t^2dt$. As a result, we get the diagonal entries of the matrix $D$ by dividing those eigenvalues we obtain above by this integral $\int_{0}^{1} \gamma_t^2dt$. Next, we follow the same procedure in the simulation to generate the log price process and calculate the pre-averaging estimator $\mathcal{B}_{m,Sp}$ and $\mathcal{B}_{m,Ws}$. Here, $p=92$, $n=23400$, $\theta=0.19$ and $\alpha=2/3$ in the simulation. Then, we compare the two matrices $\mathcal{B}_{m,Sp}$ and $\mathcal{B}_{m,Ws}$ with the original estimator $\mathcal{B}_m$.\\

\noindent
We use both the spectral norm and the Frobenius norm to measure the distance between matrices. The simulation and comparison are conducted for the 253 days in 2008 and their averaged errors are shown in Table \ref{tablea1}.
\begin{table}[H]
	\centering
	\caption{Comparison of ModelWs and ModelSp regarding the simulation error}
	\footnotesize
	\begin{tabular}{|l*{2}{c}|}			
		\hline
		Model & Spectral norm & Frobenius norm \\
		\hline
		ModelWs & $0.045$ & $0.071$ \\
		ModelSp & $0.044$ & $0.067$ \\
		\hline			
	\end{tabular}
	\label{tablea1}		
\end{table}
\noindent
For the spectral norm, the two models perform nearly the same as ModelSp only provides $2.20\%$ reduction of the error. For the Frobenius norm, ModelSp shows more improvement over the other model, reducing an average error of $4.22\%$. Since the only difference in the estimated eigenvalues of the two models is the spike part, small difference between the two models in Table \ref{tablea1} is expected. However, ModelSp shows better performance consistently. The smaller difference between the two models in term of the spectral norm can be explained by the fact that despite the absence of the explicit spiked structure, ModelWs is nevertheless capable to catch the largest eigenvalue. However, as explained previously, ModelWs is not able to estimate accurately the spikes other than the largest one. Meanwhile, the Frobenius distance takes into account all the eigenvalues. Therefore, the improvement by ModelSp over ModelWs becomes more significant in term of the Frobenius norm. This improvement is also amplified when more spikes exist in the spectrum of the ICV. \\

\noindent
In conclusion, ModelSp has a theoretical support for the empirical "bulk + spikes" structure while ModelWs does not. Additionally, ModelSp outperforms ModelWs in the statistical analysis by generating more accurate pre-averaging realized covariance matrix in the tested situations. Finally, as mentioned earlier, analogous analysis has also been conducted for Hong Kong listed stocks. The conclusions are very close to the ones presented here for the US market. Details are to be found in Section A of the supplementary document.

\section{Conclusions}
\noindent
Motivated by the "bulk + spikes" structure for the realized covariance matrix of multiple assets based on the noisy high-frequency data, and also the model in \citet{XZ15} relating the limiting spectral distribution of the pre-averaging estimator and that of the ICV under the high-dimensional setup, we incorporate the spiked model into the spectrum estimation using high-frequency intraday data in the high-dimensional setting. Consistency of the estimated spikes is proved in the main theorem. Simulation studies demonstrate the finite-sample behavior of the consistency. It is found that for various choices of spikes and their magnitudes, and different combinations of the pair of dimension and sample size, the consistency holds quite satisfactorily.\\ 

\noindent
In the real data analysis, we find that our model consistently outperforms that of \citet{XZ15}, from both empirical and statistical points of view. In addition, it is found that the magnitude of the largest spike is a potential leading indicator of the volatility of the portfolio which may be useful in practice.\\

\noindent
Several possibilities exist to further our study. First, only the consistency is established for the estimator of the spikes. One can likely establish its asymptotic normality following the approach devised in \citet{BY12}. Second, for real financial application of the spike model, one may think about various kinds of option trading strategies which make use of the change of the spikes, for example, the stability of the largest spike for the realized covariance matrix is important for dispersion trading.\\

\end{document}

\def\spacingset#1{\renewcommand{\baselinestretch}%
{#1}\small\normalsize} \spacingset{1}

%%%%%%%%%%%%%%%%%%%%%%%%%%%%%%%%%%%%%%%%%%%%%%%%%%%%%%%%%%%%%%%%%%%%%%%%%%%%%%

\if1\blind
{
  \title{\bf Supplement to "On a spiked model for large volatility matrix estimation from noisy high-frequency data"}
  \author{Keren Shen\\
  	Department of Statistics and Actuarial Science\\
  	The University of Hong Kong\\
  	Jianfeng Yao \\
  	Department of Statistics and Actuarial Science\\
  	The University of Hong Kong\\
    and \\
	Wai Keung Li \\
	Department of Statistics and Actuarial Science\\
	The University of Hong Kong\\}
  \maketitle
} \fi

\if0\blind
{
  \bigskip
  \bigskip
  \bigskip
  \begin{center}
    {\LARGE\bf Supplement to "On a spiked model for large volatility matrix estimation from noisy high-frequency data"}
\end{center}
  \medskip
} \fi

\newpage
\spacingset{1.45} % DON'T change the spacing!
\appendix

\setcounter{table}{0}
\renewcommand{\thetable}{A\arabic{table}}
\renewcommand{\thefigure}{A\arabic{figure}}
\renewcommand{\theequation}{A\arabic{equation}}
\section{The Hong Kong market}
\noindent
In this section, we compare the two models with the Hong Kong market, using the same technique in Section 4. We use $46$ stocks which are the components of the Hang Seng index. Among these $46$ stocks, 19 companies are from financial field, other companies are from conglomerates, services, consumer goods, utilities, basic materials and technology field. The whole list is available in the Appendix D. We download the tick-by-tick data from Bloomberg for transactions between 9:30 and 16:00, from September 1st, 2015 to March 10th, 2016, totally $127$ trading days. The same cleaning procedures are implemented before constructing the pre-averaging estimator $\mathcal{B}_m$. We set $\theta=0.19$ and $\alpha=2/3$.\\

\noindent
We use the same method in Section 4 to detect the number of spikes of Hong Kong data. The $127$ numbers of spikes detected are tabulated in Table \ref{table8}.\\
\begin{table}[H]
	\centering
	\caption{Distribution of the number of spikes $\hat{K}$ detected daily}
	\footnotesize
	\begin{tabular}{|l*{2}{c}|}			
		\hline
		$\hat{K}$ & Count & Percent \\
		\hline
		0 & 2 & $1.57\%$ \\
		1 & 70 & $55.12\%$ \\
		2 & 37 & $29.13\%$ \\
		3 & 17 & $13.39\%$ \\
		4 & 1 & $0.79\%$ \\
		\hline			
	\end{tabular}	
	\label{table8}	
\end{table}

\noindent
It is seen that there are one or two spikes in most trading days, which are generally less than those for US stocks. One explanation is that there are fewer categories of companies in the Hong Kong market where $40\%$ are from the financial field.\\

\begin{figure}[H]
	\centering
	\includegraphics[width=12cm]{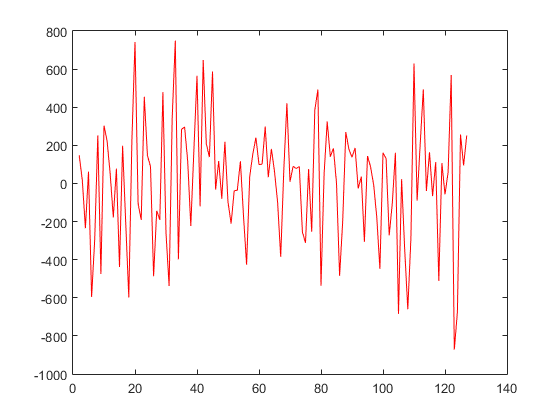}
	\caption{The daily return of the Hang Seng index}
	\label{figure13}
\end{figure}

\begin{figure}[H]
	\centering
	\includegraphics[width=12cm]{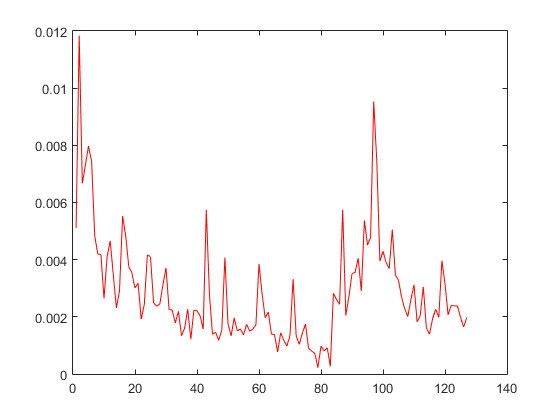}
	\caption{The largest eigenvalue of $\mathcal{B}_m$}
	\label{figure14}
\end{figure}

\noindent
Then, we plot the daily returns of the Hang Seng index and the magnitude of the largest eigenvalue for each trading day, in Figures \ref{figure13} and \ref{figure14}, respectively.

\noindent
From Figures \ref{figure13} and \ref{figure14}, we see that the first large oscillation of the largest eigenvalue of $\mathcal{B}_m$ appears in the first ten days, right before the volatile movement of the Hang Seng index, around the first twenty to forty days. Moreover, the second peak of the largest eigenvalue is around the one hundredth day, right before the second volatile oscillation of the daily return of the Hang Seng index, around 110th to 125th days. The findings are consistent with those for US stocks.\\

\begin{figure}[H]
	\centering
	\includegraphics[width=12cm]{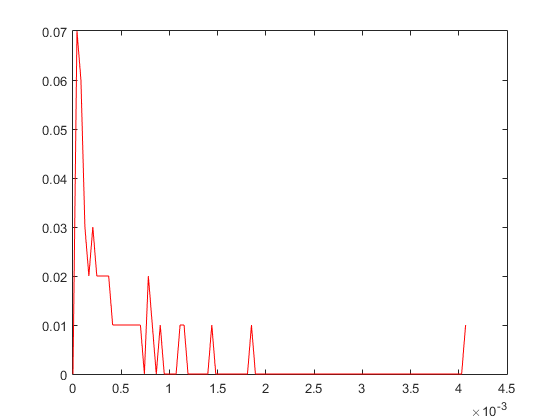}
	\caption{The ESD of $\mathcal{B}_m$ on November 11th, 2015}
	\label{figure15}
\end{figure}

\noindent
Moreover, the spectral distribution of $\mathcal{B}_m$ on November 11th, 2015 is plotted. It is seen from Figure \ref{figure15} that the "bulk + spikes" structure is present for the Hong Kong market. There is one large spike valued $4 \times 10^{-3}$, away from the other eigenvalues. In addition, the largest ten components of the first to fourth eigenvectors on the same day are presented in Table \ref{table9}.\\

\begin{table}[H]
	\centering
	\caption{Largest ten components of the first to fourth eigenvectors on November 11th, 2015}
	\scriptsize
	\begin{tabular}{|lll|lll|}			
		\hline
		Stock & Firm & Industry & Stock & Firm & Industry \\
		\hline
		\multicolumn{3}{|c|}{First eigenvector} & \multicolumn{3}{|c|}{Second eigenvector} \\
		\hline
		1088 & China Shenhua & Basic materials & 992 & Lenovo & Technology \\
		3328	& BankComm & Financial &  2319 & Mengniu Dairy & Consumer goods \\
		1299 & AIA & Financial &  1880 & Belle Int'l & Consumer goods  \\
		2628	& China Life & Financial & 3988 & Bank of China & Financial\\
		1109	& China Res Land &	Financial & 23 & Bank of E Asia & Financial \\
		883	& CNOOC  & Basic materials &  762	& China Unicom  & Technology \\
		12	& Henderson Land	& Financial & 3 & HK $\&$ China Gas & Utilities \\
		857	& PetroChina	& Basic materials & 688 & China Overseas & Financial \\
		992	& Lenovo	& Technology & 16 & SHK PPT & Financial \\
		939	& CCB & Financial & 27 & Galaxy Ent & Services \\
		\hline		
		\multicolumn{3}{|c|}{Third eigenvector} & \multicolumn{3}{|c|}{Fourth eigenvector} \\
		\hline
		322 & TingYi & Consumer goods & 2319	& Mengniu Dairy &	Consumer goods \\
		992	& Lenovo & Technology &  151 & Want Want China & Consumer goods \\
		135 & Kunlun Energy & Basic materials & 1109 & China Res Land & Financial \\
		83	& Sino Land & Financial &  2388	& BOC Hong Kong	& Financial \\
		151 & Want Want China & Consumer goods &  144 & China Mer Hold & Services \\
		2388	& BOC Hong Kong  & Financial & 823 & Link Reit & Financial \\
		4	& Wharf Holdings	& Financial & 494 & Li $\&$ Fung & Services \\
		1880 & Bell Int'l & Consumer goods & 11 & Hang Seng Bank & Financial \\
		823 & Link Reit & Financial & 23 & Bank of E Asia & Financial \\
		267	& CITIC & Basic materials & 1044 & Hengan Int'l & Consumer goods \\
		\hline		
	\end{tabular}	
	\label{table9}	
\end{table}
\noindent
All but one elements in the first eigenvector share the same sign, indicating the first eigenvector represents a market component. The other eigenvectors show no particular patterns, which include all kinds of stocks from different fields. From eigenvectors, we may conclude that we have one interpretable spike.\\

\noindent
We use the same statistical approach as in Section 4 to compare ModelSp with ModelWs. The weight accumulating procedure generates the same phenomenon of two groups of eigenvalues as that for the US stocks. There is no big difference between the largest eigenvalues of the two models while the difference is larger for other spikes. The result of the statistical comparison is shown in Table \ref{tablea2}.
\begin{table}[H]
	\centering
	\caption{Comparison of ModelWs and ModelSp regarding the simulation error}
	\footnotesize
	\begin{tabular}{|l*{2}{c}|}			
		\hline
		Model & Spectral norm & Frobenius norm \\
		\hline
		ModelWs & $0.0036$ & $0.0079$ \\
		ModelSp & $0.0035$ & $0.0078$ \\
		\hline			
	\end{tabular}
	\label{tablea2}		
\end{table}
\noindent
For the spectral norm, ModelSp produces an average error which is $3.15\%$ less than that of ModelWs. For the Frobenius norm, the error produced by ModelSp is $2.83\%$ less than that of ModelWs. In a word, ModelSp shows some improvement over ModelWs. \\

\noindent
All in all, the same conclusion can be drawn from the Hong Kong data as in the US market that ModelSp shows superior performance over ModelWs, both empirically and statistically.\\

\section{Proof of Theorem 1}
\noindent
We first set some notations that will be used throughout the proof. For any real matrix $\mathbf{B}$, $\|\mathbf{B}\| = \sqrt{\lambda_{\text{max}}(\mathbf{BB}^T)}$ denotes its spectral norm, where $\lambda_{\text{max}}$ denotes the largest eigenvalue. Moreover, $\mathbf{B}_i$ denotes the $i$th column of $\mathbf{B}$ and $b_i^j$ denotes the $j$th entry in $\mathbf{B}_i$. For any nonnegative definite matrix $\mathbf{B}$, $\mathbf{B}^{1/2}$ denotes a Hermitian square root. For any vector $\mathbf{x}$, $|\mathbf{x}|$ stands for its Euclidean norm and $x^j$ denotes its $j$th entry. Additionally, "$\stackrel{d}{=}$" means equal in distribution, $Y_n = O_p(f(n))$ means that the sequence $(|Y_n|/f(n))$ is tight, and $Y_n = o_p(f(n))$ means that $Y_n/f(n) \stackrel{p}{\rightarrow} 0$.\\ 

\noindent
Next, we detail below the set of assumptions A1-A8 borrowed from \citet{XZ15}. Note that the only difference is that in A3 we let $(\gamma_t)$ and $(\mathbf{W}_t)$ be independent, for simplicity.
\begin{enumerate}
	\item[A1.] For all $p$, $(\mathbf{X}_t)$ is a $p$-dimensional process for some drift process $\pmb{\mu}_t = (\mu_t^1, \cdots, \mu_t^p)^T$ and co-volatility process $(\mathbf{\Theta}_t)$. Moreover, almost surely, there exist $(\gamma_t) \in D([0,1];\mathbb{R})$ and $\mathbf{\Lambda}$ a $p \times p$ matrix satisfying tr$(\mathbf{\Lambda}\mathbf{\Lambda}^T)=p$ such that
	\begin{displaymath}
	\mathbf{\Theta}_t = \gamma_t\mathbf{\Lambda},
	\end{displaymath}
	where $D([0,1];\mathbb{R})$ is the space of c$\grave{a}$dl$\grave{a}$g functions from $[0,1]$ to $\mathbb{R}$.
	\item[A2.] There exists a $C_0 < \infty$ such that for all $p$ and all $j = 1, \cdots, p$, $|\mu_t^j| \le C_0$ for all $t \in [0,1)$ almost surely.
	\item[A3.] $(\gamma_t)$ and $(\mathbf{W}_t)$ are independent for all $t \in [0,1)$.
	\item[A4.] There exists a $C_1 < \infty$ such that for all $p$, $|\gamma_t| \in (1/C_1, C_1)$ for all $t \in [0,1)$ almost surely, and in addition, almost surely, $(\gamma_t)$ converges uniformly to a nonzero process $(\gamma_t^*)$ that is piecewise continuous with finitely many jumps when $p \rightarrow \infty$.
	\item[A5.] There exists a $C_2 < \infty$ such that for all $p$ and all $j = 1, \cdots, p$, the individual volatilities $\sigma_{j,t} = \sqrt{(\gamma_t)^2\cdot\sum_{k=1}^{p}(\Lambda_{jk})^2} \in (1/C_2, C_2)$ for $t \in [0,1)$ almost surely, where $\Lambda = (\Lambda_{jk})$.
	\item[A6.] For all $j = 1, \cdots, p$, the noise $(\varepsilon_i^j)$ is stationary with mean $0$ and bounded $4\ell$th moments and with $\rho$-mixing coefficients $\rho^j(r)$ satisfying\\ max$_{j=1,\cdots,p}\rho^j(r) = O(r^{-\ell})$ for some integer $\ell \ge 2$ as $r \rightarrow \infty$.
	\item[A7.] There exists a $C_3 < \infty$ and $0 \le \delta < 1/2$ such that for all $p$, $\|\text{ICV}\| \le C_3 p^{\delta}$ almost surely.
	\item[A8.] $k = \lfloor\theta n^{\alpha}\rfloor$ for some $\theta \in (0, \infty)$ and $\alpha \in [\alpha_0,1)$, and $m=\lfloor n/(2k
	)\rfloor$ satisfying that lim$_{p \rightarrow \infty}p/m=y>0$, where $\alpha_0 = \text{max}\{(3+\ell)/(2\ell+2), 2/3\}$ and $\ell$ is the integer in Assumption A6.
\end{enumerate}

\noindent
Before the proof for Theorem 1, we establish two lemmas.
\begin{lemma}
	Let
	\begin{displaymath}
	\tilde{\mathbf{V}}_i = \sum_{|j|<k} (1-\frac{|j|}{k})\int_{\frac{(2i-1)k+j-1}{n}}^{\frac{(2i-1)k+j}{n}} \pmb{\mu}_t dt = (\breve{v}_i^1, \cdots, \breve{v}_i^p)^T,
	\end{displaymath}
	\begin{displaymath}
	w_i = \sum_{|j|<k} (1-\frac{|j|}{k})^2\int_{\frac{(2i-1)k+j-1}{n}}^{\frac{(2i-1)k+j}{n}}\gamma_t^2 dt,
	\end{displaymath}
	and
	\begin{displaymath}
	\Delta \bar{\pmb{\varepsilon}}_{2i} = (\Delta \bar{\varepsilon}_{2i}^1, \cdots, \Delta \bar{\varepsilon}_{2i}^p)^T.
	\end{displaymath}
	Then, $\breve{v}_i^j = O_p(k/n)$, $w_i = O_p(k/n)$. Moreover, $\Delta \bar{\mathbf{\varepsilon}}_{2i}^j = O_p(1/\sqrt{k})$ for any $i = 1, \cdots, m$ and all $j =1, \cdots, p$.
\end{lemma}

\begin{proof}
For any process $\mathbf{V} = (\mathbf{V}_t)_{t \ge 0}$, let
\begin{displaymath}
\Delta \mathbf{V}_i = \mathbf{V}_{i/n} - \mathbf{V}_{(i-1)/n}, \bar{\mathbf{V}}_i=\frac{1}{k}\sum_{j=0}^{k-1}\mathbf{V}_{((i-1)k+j)/n}, \text{ and } \Delta \bar{\mathbf{V}}_{2i} = \bar{\mathbf{V}}_{2i} - \bar{\mathbf{V}}_{2i-1}.
\end{displaymath}
Note that $\Delta \bar{\mathbf{V}}_{2i}$ can be written as
\begin{eqnarray}
\Delta \bar{\mathbf{V}}_{2i} &=& \frac{1}{k}\sum_{j=0}^{k-1}(\mathbf{V}_{((2i-1)k+j)/n} - \mathbf{V}_{((2i-2)k+j)/n}) \nonumber \\
&=& \frac{1}{k}\sum_{j=0}^{k-1}\sum_{\ell=1}^{k} \Delta \mathbf{V}_{(2i-2)k+j+\ell} \nonumber \\
&=& \sum_{|j|<k}(1-\frac{|j|}{k}) \Delta \mathbf{V}_{(2i-2)k+j}. \nonumber
\end{eqnarray}
The result is then easily proved by Assumptions A2, A4, and A6.
\end{proof}	

\begin{lemma}
	Let $\breve{\mathbf{\Sigma}}^{1/2} \mathbf{Z} := (\breve{Z}^1, \cdots, \breve{Z}^p)^T$, where $\mathbf{Z}$ is a $p$-dimensional vector consisting of i.i.d standard normal random variables. Then, $\breve{Z}^j = O_p(1)$ for all $j = 1,\cdots, p$.
\end{lemma}

\begin{proof}
	Let $\breve{Z}^j = \sum_{k=1}^{p} \lambda_k^jZ^k$, where $Z^k$'s are i.i.d standard normal distributed variates. Therefore, from Assumption A5,
	\begin{displaymath}
	\breve{Z}^j = \sum_{k=1}^{p} \lambda_k^jZ^k \stackrel{d}{=} N(0, \sum_{k=1}^{p}(\lambda_k^j)^2) = O_p(1)
	\end{displaymath} 
	for all $p$ and $j$.
\end{proof}

\begin{proof}[Proof of Theorem 1]
	We have
	\begin{eqnarray}
	\Delta\bar{\mathbf{Y}}_{2i} &=& \Delta\bar{\mathbf{X}}_{2i} + \Delta\bar{\pmb{\varepsilon}}_{2i} \nonumber \\
	&=& \tilde{\mathbf{V}}_i + \sqrt{w_i}\breve{\mathbf{\Sigma}}^{1/2} \mathbf{Z}_i + \Delta\bar{\pmb{\varepsilon}}_{2i}, \nonumber 
	\end{eqnarray}
	where
	\begin{displaymath}
	\tilde{\mathbf{V}}_i = \sum_{|j|<k} (1-\frac{|j|}{k})\int_{\frac{(2i-1)k+j-1}{n}}^{\frac{(2i-1)k+j}{n}} \pmb{\mu}_t dt,
	\end{displaymath}
	\begin{displaymath}
	w_i = \sum_{|j|<k} (1-\frac{|j|}{k})^2\int_{\frac{(2i-1)k+j-1}{n}}^{\frac{(2i-1)k+j}{n}}\gamma_t^2 dt,
	\end{displaymath}
	and $\mathbf{Z}_i = (Z_i^1, \cdots, Z_i^p)^T$ consists of independent standard normal random variates. As a result,
	\begin{eqnarray}
	\tilde{\mathbf{\Sigma}} &=& \frac{p}{m} \sum_{i=1}^{m}\frac{\Delta\bar{\mathbf{Y}}_{2i}(\Delta\bar{\mathbf{Y}}_{2i})^T}{|\Delta\bar{\mathbf{Y}}_{2i}|^2} \nonumber \\
	&=& \frac{p}{m} \sum_{i=1}^{m}\frac{(\tilde{\mathbf{V}}_i + \sqrt{w_i}\breve{\mathbf{\Sigma}}^{1/2} \mathbf{Z}_i + \Delta\bar{\pmb{\varepsilon}}_{2i})(\tilde{\mathbf{V}}_i + \sqrt{w_i}\breve{\mathbf{\Sigma}}^{1/2} \mathbf{Z}_i + \Delta\bar{\pmb{\varepsilon}}_{2i})^T}{|\tilde{\mathbf{V}}_i + \sqrt{w_i}\breve{\mathbf{\Sigma}}^{1/2} \mathbf{Z}_i + \Delta\bar{\pmb{\varepsilon}}_{2i}|^2} \nonumber \\
	&=& \frac{p}{m} \sum_{i=1}^{m}\frac{(\frac{\tilde{\mathbf{V}}_i + \Delta\bar{\pmb{\varepsilon}}_{2i}}{\sqrt{w_i}} + \breve{\mathbf{\Sigma}}^{1/2} \mathbf{Z}_i)(\frac{\tilde{\mathbf{V}}_i + \Delta\bar{\pmb{\varepsilon}}_{2i}}{\sqrt{w_i}} + \breve{\mathbf{\Sigma}}^{1/2} \mathbf{Z}_i)^T}{|\frac{\tilde{\mathbf{V}}_i + \Delta\bar{\pmb{\varepsilon}}_{2i}}{\sqrt{w_i}} + \breve{\mathbf{\Sigma}}^{1/2} \mathbf{Z}_i|^2} \nonumber \\
	&=& \frac{p}{m} \sum_{i=1}^{m}\frac{(\mathbf{V}_i + \breve{\mathbf{\Sigma}}^{1/2} \mathbf{Z}_i)(\mathbf{V}_i + \breve{\mathbf{\Sigma}}^{1/2} \mathbf{Z}_i)^T}{|\mathbf{V}_i + \breve{\mathbf{\Sigma}}^{1/2} \mathbf{Z}_i|^2}, \nonumber
	\end{eqnarray}
	where $\mathbf{V}_i = (\tilde{\mathbf{V}}_i + \Delta\bar{\pmb{\varepsilon}}_{2i})/\sqrt{w_i}$. From Lemma 1, we know that $\tilde{v}_i^j = O(k/n)$, $\Delta\bar{\pmb{\varepsilon}}_{2i}^j = O_p(1/\sqrt{k})$, and $w_i = O_p(k/n)$, for all $i$ and $j$. Therefore, from Assumption A8, 
	\begin{equation}
	v_i^j = O_p(\sqrt{k/n}) = O_p(1/\sqrt{p}),
	\end{equation}
	so that $|\mathbf{V}_i|$ are uniformly bounded. \\
	
	\noindent
	We borrow the following convergence result proved in \citet{XZ15}:
	\begin{displaymath}
	\text{max}_{i=1,\cdots,m}||\breve{\mathbf{\Sigma}}^{1/2} \mathbf{Z}_i|^2/p - 1| = \text{max}_{i=1,\cdots,m}|\mathbf{Z}_i^T \breve{\mathbf{\Sigma}} \mathbf{Z}_i/p - 1| \rightarrow 0, \text{ almost surely.}
	\end{displaymath}
	\noindent
	Consequently, we have
	\begin{displaymath}
	\text{max}_{i=1,\cdots,m}||\mathbf{V}_i + \breve{\mathbf{\Sigma}}^{1/2} \mathbf{Z}_i|^2/p - 1| \rightarrow 0, \quad \text{ almost surely.}
	\end{displaymath}
	In other words, we may focus on 
	\begin{eqnarray}
	\mathbf{S}_m &=& \frac{1}{m} \sum_{i=1}^{m}(\mathbf{V}_i + \breve{\mathbf{\Sigma}}^{1/2} \mathbf{Z}_i)(\mathbf{V}_i + \breve{\mathbf{\Sigma}}^{1/2} \mathbf{Z}_i)^T \nonumber \\
	&=&  \frac{1}{m} \sum_{i=1}^{m}(\mathbf{V}_i + \mathbf{W}_i)(\mathbf{V}_i + \mathbf{W}_i)^T, \nonumber
	\end{eqnarray}
	where we let $\mathbf{W}_i := \breve{\mathbf{\Sigma}}^{1/2} \mathbf{Z}_i$.\\
	
	\noindent
	We have
	\begin{displaymath}
	\breve{\mathbf{\Sigma}} = \begin{pmatrix}
	\mathbf{V}_s & \mathbf{0} \\
	\mathbf{0} & \mathbf{V}_p  
	\end{pmatrix}
	,
	\end{displaymath}
	where $\mathbf{V}_s$ is of size $K \times K$ and $\mathbf{V}_p$ is of size $(p-K) \times (p-K)$. Similar to the approach of \citet{YBZ15}, decompose the $p$-dimensional vectors $\mathbf{V}_i$, $\mathbf{W}_i$ and $\mathbf{Z}_i$ into blocks of size $K$ and $p-K$, respectively:
	\begin{displaymath}
	\mathbf{V}_i = \begin{pmatrix}
	\mathbf{V}_{1i} \\
	\mathbf{V}_{2i}
	\end{pmatrix}
	, \quad
	\mathbf{W}_i = \begin{pmatrix}
	\mathbf{W}_{1i} \\
	\mathbf{W}_{2i}
	\end{pmatrix}
	, \quad
	\mathbf{Z}_i = \begin{pmatrix}
	\mathbf{Z}_{1i} \\
	\mathbf{Z}_{2i}
	\end{pmatrix}
	.
	\end{displaymath} 
	Then, we may write
	\begin{eqnarray}
	\mathbf{S}_m &=& \frac{1}{m}\sum_{i=1}^{m} (\mathbf{V}_i+\mathbf{W}_i)(\mathbf{V}_i+\mathbf{W}_i)^T \nonumber \\
	&=& \frac{1}{m} \begin{pmatrix}
	(\mathbf{V}_{(1)}+\mathbf{W}_{(1)})(\mathbf{V}_{(1)}+\mathbf{W}_{(1)})^T & (\mathbf{V}_{(1)}+\mathbf{W}_{(1)})(\mathbf{V}_{(2)}+\mathbf{W}_{(2)})^T \\
	(\mathbf{V}_{(2)}+\mathbf{W}_{(2)})(\mathbf{V}_{(1)}+\mathbf{W}_{(1)})^T & (\mathbf{V}_{(2)}+\mathbf{W}_{(2)})(\mathbf{V}_{(2)}+\mathbf{W}_{(2)})^T  
	\end{pmatrix} \nonumber \\
	&:=& \begin{pmatrix}
	\mathbf{S}_{11} & \mathbf{S}_{12} \\
	\mathbf{S}_{21} & \mathbf{S}_{22} 
	\end{pmatrix}
	, \nonumber
	\end{eqnarray} 
	where 
	\begin{displaymath}
	\mathbf{V}_{(1)} = (\mathbf{V}_{11}, \cdots, \mathbf{V}_{1m}), \quad \mathbf{V}_{(2)} = (\mathbf{V}_{21}, \cdots, \mathbf{V}_{2m}),
	\end{displaymath} 
	and  
	\begin{displaymath}
	\mathbf{W}_{(1)} = (\mathbf{W}_{11}, \cdots, \mathbf{W}_{1m}), \quad \mathbf{W}_{(2)} = (\mathbf{W}_{21}, \cdots, \mathbf{W}_{2m}).
	\end{displaymath}  
	Similarly, define the analogous decomposition for $\mathbf{Z}_i$ vectors to form $\mathbf{Z}_{(1)}$ and $\mathbf{Z}_{(2)}$ such that
	\begin{displaymath}
	\mathbf{W}_{(1)} = \mathbf{V}_s^{1/2}\mathbf{Z}_{(1)}, \quad \mathbf{W}_{(2)} = \mathbf{V}_p^{1/2}\mathbf{Z}_{(2)}.
	\end{displaymath} 
	An eigenvalue $\lambda_i$ of $\mathbf{S}_m$ that is not an eigenvalue of $\mathbf{S}_{22}$ satisfies
	\begin{displaymath}
	0 = |\lambda_i\mathbf{I}_p - \mathbf{S}_m| = |\lambda_i\mathbf{I}_{p-K} - \mathbf{S}_{22}|\cdot|\lambda_i\mathbf{I}_{K} - \mathbf{K}_m(\lambda_i)|,
	\end{displaymath} 
	where 
	\begin{displaymath}
	\mathbf{K}_m(\ell) = \mathbf{S}_{11} + \mathbf{S}_{12}(\ell\mathbf{I}_{p-K}-\mathbf{S}_{22})^{-1}\mathbf{S}_{21}.
	\end{displaymath}
	This will happen if $\lambda_i$ tends to a location outside the support of the limiting spectral distribution (LSD) of $\mathbf{S}_m$ (the same as the LSD of $\mathbf{S}_{22}$) so that for such $\lambda_i$ and large $m$, it will hold that $|\lambda_i\mathbf{I}_{p-K} - \mathbf{S}_{22}|\ne 0$, so that necessarily
	\begin{displaymath}
	|\lambda_i\mathbf{I}_{K} - \mathbf{K}_m(\lambda_i)| = 0.
	\end{displaymath}
	Consider a real number $\ell$ outside the support of the LSD of $\mathbf{S}_{22}$, it holds that
	\begin{eqnarray}
	\mathbf{K}_m(\ell) &=& \mathbf{S}_{11} + \mathbf{S}_{12}(\ell\mathbf{I}_{p-K}-\mathbf{S}_{22})^{-1}\mathbf{S}_{21} \nonumber \\
	&=& \frac{1}{m}(\mathbf{V}_{(1)} + \mathbf{W}_{(1)})\{\mathbf{I}_m + \frac{1}{m}(\mathbf{V}_{(2)} + \mathbf{W}_{(2)})^T(\ell\mathbf{I}_{p-K}-\frac{1}{m}(\mathbf{V}_{(2)} + \mathbf{W}_{(2)})(\mathbf{V}_{(2)} + \mathbf{W}_{(2)})^T)^{-1} \nonumber \\
	&& (\mathbf{V}_{(2)} + \mathbf{W}_{(2)})\}(\mathbf{V}_{(1)} + \mathbf{W}_{(1)})^T \nonumber \\
	&=& \frac{\ell}{m} (\mathbf{V}_{(1)} + \mathbf{W}_{(1)})(\ell\mathbf{I}_m - \frac{1}{m}(\mathbf{V}_{(2)} + \mathbf{W}_{(2)})^T(\mathbf{V}_{(2)} + \mathbf{W}_{(2)}))^{-1}(\mathbf{V}_{(1)} + \mathbf{W}_{(1)})^T \nonumber \\
	&=& \frac{\ell}{m} (\mathbf{V}_{(1)} + \mathbf{V}_s^{1/2}\mathbf{Z}_{(1)})(\ell\mathbf{I}_m - \frac{1}{m}(\mathbf{V}_{(2)} + \mathbf{W}_{(2)})^T(\mathbf{V}_{(2)} + \mathbf{W}_{(2)}))^{-1}(\mathbf{V}_{(1)}^T + \mathbf{Z}_{(1)}^T\mathbf{V}_s^{1/2}), \nonumber
	\end{eqnarray} 
	where the third equality uses the following identity: for $\ell \ne 0$ which is not an eigenvalue of $\mathbf{A}^T\mathbf{A}$,
	\begin{displaymath}
	\mathbf{I}_m + \mathbf{A}(\ell\mathbf{I}_{p-K}-\mathbf{A}^T\mathbf{A})^{-1}\mathbf{A}^T \equiv \ell(\ell\mathbf{I}_m - \mathbf{A}\mathbf{A}^T)^{-1}.
	\end{displaymath}
	Since $\ell$ is outside the support of the LSD of $\mathbf{S}_{22}$, for large $m$, the norm of $(\ell\mathbf{I}_m - \frac{1}{m}(\mathbf{V}_{(2)} + \mathbf{W}_{(2)})^T(\mathbf{V}_{(2)} + \mathbf{W}_{(2)}))^{-1}$ is bounded. Therefore,
	\begin{eqnarray}
	&& \|\frac{\ell}{m} \mathbf{V}_{(1)}(\ell\mathbf{I}_m - \frac{1}{m}(\mathbf{V}_{(2)} + \mathbf{W}_{(2)})^T(\mathbf{V}_{(2)} + \mathbf{W}_{(2)}))^{-1}\mathbf{V}_{(1)}^T \| \nonumber\\
	&\le& \frac{\ell}{m}\|\mathbf{V}_{(1)}\|^2 \|(\ell\mathbf{I}_m - \frac{1}{m}(\mathbf{V}_{(2)} + \mathbf{W}_{(2)})^T(\mathbf{V}_{(2)} + \mathbf{W}_{(2)}))^{-1}\| \nonumber \\
	&=& \frac{C}{m} \|\mathbf{V}_{(1)}\mathbf{V}_{(1)}^T\| \nonumber \\
	&\le& \frac{C}{m} \text{tr}(\mathbf{V}_{(1)}\mathbf{V}_{(1)}^T) \nonumber \\
	&=&  \frac{C}{m} \sum_{i=1}^{m} \sum_{j=1}^{K} (v_i^j)^2 \nonumber \\
	&=& O_p(\frac{1}{p}) = o_p(1), \nonumber
	\end{eqnarray}
	as max$_{i= 1, \cdots,m}v_i^j = O_p(1/\sqrt{p})$ for all $p$ and all $j$, from Equation (A1). From Lemma $2$, we know that the order of the magnitude of the elements in $\mathbf{Z}_{(1)}^T\mathbf{V}_s^{1/2}$ is $O_p(1)$ so that similarly we have
	\begin{eqnarray}
	&& \|\frac{\ell}{m} \mathbf{V}_{(1)}(\ell\mathbf{I}_m - \frac{1}{m}(\mathbf{V}_{(2)} + \mathbf{W}_{(2)})^T(\mathbf{V}_{(2)} + \mathbf{W}_{(2)}))^{-1}\mathbf{Z}_{(1)}^T\mathbf{V}_s^{1/2} \| \nonumber\\
	&\le& \frac{\ell}{m}\|\mathbf{V}_{(1)}\| \|(\ell\mathbf{I}_m - \frac{1}{m}(\mathbf{V}_{(2)} + \mathbf{W}_{(2)})^T(\mathbf{V}_{(2)} + \mathbf{W}_{(2)}))^{-1}\|\|\mathbf{Z}_{(1)}^T\mathbf{V}_s^{1/2}\| \nonumber \\
	&\le& \frac{C}{m} \sqrt{m} = O_p(\frac{1}{\sqrt{p}}) = o_p(1). \nonumber
	\end{eqnarray}
	Therefore, by the Law of Large Numbers, 
	\begin{eqnarray}
	\mathbf{K}_m(\ell) &=& \mathbf{V}_s\cdot [\ell \cdot  \text{tr}(\ell\mathbf{I}_m - \frac{1}{m}(\mathbf{V}_{(2)} + \mathbf{W}_{(2)})^T(\mathbf{V}_{(2)} + \mathbf{W}_{(2)}))^{-1}/m] + o_{a.s.}(1) \nonumber \\
	&=& - \mathbf{V}_s\cdot \ell \underline{\breve{s}}(\ell) + o_{a.s.}(1), \nonumber
	\end{eqnarray}
	where $\underline{\breve{s}}(\cdot)$ is the Stieltjes transform of the LSD of $\frac{1}{m}(\mathbf{V}_{(2)} + \mathbf{W}_{(2)})^T(\mathbf{V}_{(2)} + \mathbf{W}_{(2)})$. From \citet{XZ15}, the LSD of $\frac{1}{m}(\mathbf{V}_{(2)} + \mathbf{W}_{(2)})^T(\mathbf{V}_{(2)} + \mathbf{W}_{(2)})$ and $\frac{1}{m}\mathbf{W}_{(2)}^T\mathbf{W}_{(2)}$ should be the same, so that 
	\begin{displaymath}
	\mathbf{K}_m(\ell) = - \mathbf{V}_s\cdot \ell \underline{s}(\ell) + o_{a.s.}(1)
	\end{displaymath}
	where $\underline{s}(\cdot)$ is the Stieltjes transform of the LSD of $\frac{1}{m}\mathbf{W}_{(2)}^T\mathbf{W}_{(2)}$.\\
	
	\noindent
	Therefore, almost surely, $\ell$ is an eigenvalue of $- \mathbf{V}_s\cdot \ell \underline{s}(\ell)$ so that
	\begin{displaymath}
	\underline{s}(\ell) = -1/\breve{\alpha}_j.
	\end{displaymath}
	As 
	\begin{displaymath}
	\breve{\psi}(\alpha) = \alpha + y\int \frac{t\alpha}{\alpha-t} d\breve{H}(t),
	\end{displaymath}
	which is the functional inverse of the function $x \mapsto -1/\underline{s}(x)$, and $\breve{\psi}(\cdot)$ is well defined for all $\alpha \notin \Gamma_{\breve{H}}$. As a result, if such $\ell$ exists, $\ell$ must satisfy the equation
	\begin{displaymath}
	\ell = \breve{\psi}(\breve{\alpha}_j)
	\end{displaymath}
	for some $\breve{\alpha}_j$. Furthermore, with a similar argument as in Theorems 4.1 and 4.2 of \citet{SC95}, we know that $\ell = \breve{\psi}(\breve{\alpha})$ is outside the support of the LSD $F_{y,\breve{H}}$ if and only if $\breve{\psi}'(\breve{\alpha})>0$.\\
	
	\noindent
	In addition, from the proof of Lemma A.1 in \citet{XZ15},
	\begin{displaymath}
	\text{lim}_{p\rightarrow \infty} 3\frac{\sum_{i=1}^{m}|\Delta\bar{\mathbf{Y}}_{2i}|^2}{p} = \text{lim}_{p\rightarrow \infty} 3\frac{\sum_{i=1}^{m}|\Delta\bar{\mathbf{X}}_{2i}|^2}{p} = \zeta,
	\end{displaymath}
	where $\zeta = \int_{0}^{1} (\gamma_t^*)^2 dt$. Then, it is easily shown that for a spiked eigenvalue $\alpha_j$ of the ICV satisfying $\psi'(\alpha_j) > 0$, there is an eigenvalue $\lambda_j$ of $\mathcal{B}_m$ such that
	\begin{displaymath}
	\lambda_j \stackrel{a.s.}{\rightarrow} \psi_j = \psi(\alpha_j),
	\end{displaymath}
	where 
	\begin{displaymath}
	\psi(\alpha) = \alpha + y\int \frac{t\alpha}{\alpha-t} dH(t),
	\end{displaymath}
	and 
	\begin{displaymath}
	H(x) = \breve{H}(x/\zeta) \text{ for all }x \ge 0.
	\end{displaymath}
\end{proof}

\newpage
\section{Details of 92 stocks used for the US market}
\begin{table}[H]
	\centering
	\caption{List of 92 stocks in the S$\&$P 100 index}
	\scriptsize
	\begin{tabular}{|lll|lll|}			
		\hline
		Stock & Firm & Industry & Stock & Firm & Industry \\
		\hline
		AAPL &	Apple Inc &	Consumer goods & HON &	Honeywell &	Industrial goods \\
		ABT	& Abbott Laboratories &	Healthcare &   IBM &	International Business Machines &	Technology\\
		ACN	& Accenture plc	& Technology &   INTC &	Intel Corp &	Technology\\
		AIG	& American International Group &	Financial & JNJ	& Johnson $\&$ Johnson Inc	& Healthcare\\
		ALL	& Allstate Corp &	Financial &  JPM	& JP Morgan Chase $\&$ Co &	Financial\\
		AMGN &	Amgen Inc &	Healthcare &  KO &	The Coca-Cola Co &	Consumer goods\\
		AMZN &	Amazon.com &	Services &   LLY &	Eli Lilly and Co &	Healthcare\\
		APA	& Apache Corp &	Basic materials &  LMT &	Lockheed-Martin &	Industrial goods\\
		APC	& Anadarko Petroleum Corp &	Basic materials &  LOW &	Lowe's &	Services\\
		AXP	& American Express Inc &	financial &  MA &	Mastercard Inc &	Financial\\
		BA	& Boeing Co &	Industrial goods &  MCD &	McDonald's Corp &	Services\\
		BAC	& Bank of America Corp &	Financial &  MDT &	Medtronic Inc &	Healthcare\\
		BAX	& Baxter International Inc &	Healthcare &  MET &	Metlife Inc &	Financial\\
		BIIB &	Biogen Idec	& Healthcare  & MMM &	3M Company &	Industrial goods\\
		BK	& Bank of New York &	Financial &  MO &	Altria Group &	Consumer goods\\
		BLK	& BlackRock Inc	& Financial &  MON &	Monsanto & 	Basic materials\\
		BMY	& Bristol-Myers Squibb &	Healthcare &  MRK &	Merck $\&$ Co &	Healthcare\\
		C	& Citigroup Inc	& Financial &  MS &	Morgan Stanley &	Financial\\
		CAT	& Caterpillar Inc &	Industrial goods &  MSFT &	Microsoft &	Technology\\
		CL	& Colgate-Palmolive &	Consumer goods &  NKE &	Nike &	Consumer goods\\
		CMCSA &	Comcast Corp &	Services &  NOV &	National Oilwell Varco &	Basic materials\\
		COF	& Capital One Financial Corp &	Financial &  NSC &	Norfolk Southern Corp &	Services\\
		COP	& ConocoPhillips &	Basic materials &  ORCL &	Oracle Corp &	Technology\\
		COST &	Costco &	Services &  OXY &	Occidental Petroleum Corp &	Basic materials\\
		CSCO &	Cisco Systems &	Technology &  PEP &	Pepsico Inc &	Consumer goods\\
		CVS	& CVS Caremark &	Healthcare &  PFE &	Pfizer Inc &	Healthcare\\
		CVX	& Chevron &	Basic materials  & PG &	Procter $\&$ Gamble Co &	Consumer goods\\
		DD	& DuPont &	Basic materials  & QCOM &	Qualcomm Inc &	technology\\
		DIS	& Walt Disney &	Services &  RTN &	Raytheon Co (NEW) &	Industrial goods\\
		DOW	& Dow Chemical &	Basic materials &  SBUX &	Starbucks Corp &	Services\\
		DVN	& Devon Energy &	Basic materials &  SLB &	Schlumberger &	Basic materials\\
		EBAY &	eBay Inc &	Services  & SO &	Southern Company &	Utilities\\
		EMC	& EMC Corp &	Technology &  SPG &	Simon Property Group &	Financial\\
		EMR	& Emerson Electric &	Industrial goods &  T &	AT$\&$T Inc &	Technology\\
		EXC	& Exelon &	Utilities &  TGT &	Target Corp &	Services\\
		F	& Ford Motor &	Consumer goods &  TWX &	Time Warner Inc &	Services\\
		FCX	& Freeport-McMoran &	Basic materials &  TXN &	Texas Instruments &	Technology\\
		FDX	& FedEx	& Services &  UNH &	UnitedHealth Group Inc &	Healthcare\\
		GD	& General Dynamics &	Industrial goods &  UNP &	Union Pacific Corp &	Services\\
		GE	& General Electric &	Industrial goods &  UPS &	United Parcel Service Inc &	Services\\
		GILD &	Gilead Sciences &	Healthcare &  USB &	US Bancorp &	Financial\\
		GM	& General Motors &	Consumer goods &  UTX &	United Technologies Corp &	Industrial goods\\
		GOOG &	Google Inc &	Technology  & VZ &	Verizon Communications Inc &	Technology\\
		GS	& Goldman Sachs &	Financial &  WFC &	Wells Fargo	& Financial\\
		HAL	& Halliburton &	Basic materials &  WMT &	Wal-Mart &	Services\\
		HD	& Home Depot &	Services  & XOM	 & Exxon Mobil Corp	& Basic materials\\
		\hline		
	\end{tabular}		
\end{table}

\newpage
\section{Details of 46 stocks used for the Hong Kong market}
\begin{table}[H]
	\centering
	\caption{List of 46 stocks in the Hang Seng index}
	\footnotesize
	\begin{tabular}{|lll|lll|}			
		\hline
		Stock Code & Firm & Industry & Stock Code & Firm & Industry\\
		\hline
		1 &	CKH Holdings &	Conglomerates & 267 &	CITIC &	Basic materials\\
		101 &	Hang Lung PPT &	Conglomerates & 27 &	Galaxy Ent &	Services\\
		1044 &	Hengan Int'l &	Consumer goods & 291 &	China Res Beer &	Conglomerates\\
		1088 & 	China Shenhua &	Basic materials & 293 &	Cathay Pac Air &	Services\\
		11 &	Hang Seng Bank &	Financial & 3 &	HK $\&$ China Gas &	Utilities\\
		1109 &	China Res Land &	Financial & 322 &	TingYi &	Consumer goods\\
		1113 &	CK Property &	Conglomerates & 3328 &	BankComm &	Financial\\
		12 &	Henderson Land &	Financial & 386 &	Sinopec Corp &	Basic materials\\
		1299 &	AIA &	Financial & 388 &	HKEX &	Financial\\
		135 &	Kunlun Energy &	Basic materials & 3988 &	Bank of China &	Financial\\
		1398 &	ICBC &	Financial & 4 &	Wharf Holdings &	Financial\\
		144 &	China Mer Hold &	Services & 494 &	Li $\&$ Fung &	Services\\
		151 &	Want Want China &	Consumer goods & 5 &	HSBC Holdings &	Financial\\
		16 &	SHK PPT &	Financial & 688 &	China Overseas &	Financial\\
		1880 &	Belle Int'l &	Consumer goods & 762 &	China Unicom &	Technology\\
		19 &	Swire Pacific &	Conglomerates & 823 & 	Link Reit &	Financial\\
		1928 &	Sands China Ltd &	Services & 83 &	Sino Land &	Financial\\
		2 &	CLP Holdings &	Utilities & 836 &	China Res Power &	Utilities\\
		23 &	Bank of E Asia &	Financial & 857 &	PetroChina &	Basic materials\\
		2318 &	Ping An &	Financial & 883 &	CNOOC &	Basic materials\\
		2319 &	Mengniu Dairy &	Consumer goods & 939 &	CCB & 	Financial\\
		2388 &	BOC Hong Kong &	Financial & 941 &	China Mobile &	Technology\\
		2628 &	China Life &	Financial &	992 &	Lenovo &	Technology\\
		\hline		
	\end{tabular}		
\end{table}